\documentclass[%
 reprint,
 amsmath,amssymb,
 aps,
]{revtex4-1}

\usepackage{comment}
\usepackage{color}
\usepackage{mathrsfs} 
\usepackage{xspace}	
\usepackage{graphicx}
\usepackage{dcolumn}
\usepackage{bm}
\usepackage{hyperref}
\usepackage[mathlines]{lineno}
\relax 

\begin{document}

\newcommand{\pt}{\mbox{$p_T$}\xspace}
\newcommand{\Npart}{\mbox{$N_{\rm part}$}\xspace}
\newcommand{\Ncoll}{\mbox{$N_{\rm coll}$}\xspace}
\newcommand{\Nch}{\mbox{$N_{\rm ch}$}\xspace}
\newcommand{\sqs}{\mbox{$\sqrt{s}$}\xspace}
\newcommand{\sqsn}{\mbox{$\sqrt{s_{_{NN}}}$}\xspace}
\newcommand{\dau}{\mbox{$d$$+$Au}\xspace}
\newcommand{\pau}{\mbox{$p$$+$Au}\xspace}
\newcommand{\auau}{\mbox{Au$+$Au}\xspace}
\newcommand{\pal}{\mbox{$p$$+$Al}\xspace}
\newcommand{\hau}{\mbox{$^3$He$+$Au}\xspace}
\newcommand{\pp}{\mbox{$p$$+$$p$}\xspace}
\newcommand{\ppb}{\mbox{$p$$+$Pb}\xspace}
\newcommand{\pbpb}{\mbox{Pb$+$Pb}\xspace}
\newcommand{\pa}{\mbox{$p$$+$$A$}\xspace}
\newcommand{\nucnuc}{\mbox{$A$$+$$A$}\xspace}
\newcommand{\da}{\mbox{$d$$+$$A$}\xspace}
\newcommand{\pdhau}{\mbox{$p/d/^{3}$He$+$Au}\xspace}
\newcommand{\pdha}{\mbox{$p/d/^{3}$He$+$A}\xspace}
\newcommand{\ampt}{{\sc ampt}\xspace}
\newcommand{\sonic}{{\sc sonic}\xspace}
\newcommand{\ipjazma}{{\sc IP-Jazma}\xspace}
\newcommand{\Qsp}{\mbox{$Q_s(\mathrm{proj})$}\xspace}
\newcommand{\Qst}{\mbox{$Q_s(\mathrm{targ})$}\xspace}
\newcommand{\QspSq}{\mbox{$Q_s^2(\mathrm{proj})$}\xspace}
\newcommand{\QstSq}{\mbox{$Q_s^2(\mathrm{targ})$}\xspace}

\preprint{APS/123-QED}

\title{Assessing saturation physics explanations of collectivity in small collision systems with the \ipjazma model}

\author{J.L. Nagle}
 \email{jamie.nagle@colorado.edu}
 \affiliation{University of Colorado Boulder and CEA/IPhT/Saclay}
 
\author{W.A. Zajc}%
 \email{zajc@nevis.columbia.edu}
\affiliation{Columbia University}

\date{\today}

\begin{abstract}
Experimental measurements in relativistic collisions of small systems from \pp to \pdha at the Relativistic Heavy Ion Collider (RHIC) and the Large Hadron Collider (LHC) reveal particle emission patterns that are strikingly similar to those observed in \nucnuc collisions of large nuclei.   One explanation of these patterns is the formation of small droplets of quark-gluon plasma (QGP) followed by hydrodynamic evolution.    A geometry engineering program was proposed~\cite{Nagle:2013lja} to further investigate these emission patterns, and the experimental data from that program in \pau, \dau, \hau collisions for elliptic and triangular anisotropy coefficients $v_{2}$ and 
$v_{3}$ follow the  pattern predicted by hydrodynamic calculations~\cite{Aidala:2018mcw}.  One alternative approach, referred to as initial-state correlations, suggests that for small systems the patterns observed in the final-state hadrons are encoded at the earliest moments of the collision, and therefore require no final-state parton scattering or hydrodynamic evolution~\cite{Dusling:2012iga,Dusling:2013qoz}.  Recently, new calculations 
using only initial-state correlations, in the dilute-dense approximation of gluon saturation physics, reported striking agreement with the $v_2$ patterns observed in \pdhau data at RHIC~\cite{Mace:2018vwq}.   The reported results are counterintuitive and thus we aim here to reproduce some of the basic features of these calculations.  
In this first investigation, we  provide a description of our model, \ipjazma, and investigate its implications for saturation scales, multiplicity distributions and eccentricities, reserving for later work the analysis of momentum spectra and azimuthal anisotropies. We find that our implementation of the saturation physics model reproduces the results of the MSTV calculation of the multiplicity distribution in \dau collisions at RHIC. However, our investigations, together with existing data, 
call into question some of the essential elements reported in Ref.~\cite{Mace:2018vwq}.
\end{abstract}

\pacs{Valid PACS appear here}
\maketitle


\section{Introduction}
The standard model for the evolution of the medium in heavy ion (\nucnuc) collisions at RHIC and the LHC assumes the matter proceeds through a quark-gluon plasma stage described quantitatively via nearly inviscid hydrodynamics~\cite{Heinz:2013th}.    
Observations in \pp and \pdha collisions of features similar to those found
in \nucnuc collisions raise the question of whether one forms quark-gluon plasma in these smaller systems as well, albeit in a smaller volume and evolving for a shorter lifetime -- for recent reviews see Refs.~\cite{Nagle:2018nvi,Romatschke:2017ejr,Dusling:2015gta}.   A particularly striking theoretical calculation, with the evocative title {\em One Fluid to Rule Them All}~\cite{Weller:2017tsr} is the simultaneous matching of viscous hydrodynamic calculations with \pp, \ppb, and \pbpb data from the LHC using a common set of initial conditions and hydrodynamic input parameters.
A geometry engineering program was
proposed at RHIC specifically to test the hypothesis that the initial geometry was
responsible for the momentum anisotropies by
generating droplets with different magnitudes of ellipticity and triangularity via \pau, \dau, and \hau collisions~\cite{Nagle:2013lja}.    
This program led to experimental measurements by the PHENIX experiment at RHIC~\cite{Aidala:2018mcw} that are found to be in good quantitative agreement with the hydrodynamic predictions.

Since these findings have large impact, it is scientifically mandated to scrutinize the hydrodynamic calculations and their sensitivities to various inputs, while at the same time to fully explore alternative explanations.   One such alternative explanation was proposed shortly after
the first collective-type signatures were observed in high-multiplicity \pp collisions at the LHC~\cite{Khachatryan:2010gv}.   The calculation is done in the context of gluon saturation physics and finds azimuthal correlations between particles that extend over large rapidity ranges~\cite{Dusling:2012iga,Dusling:2013qoz}.    Many additional papers have followed within this saturation
physics framework where the correlations are generated in the initial state, just at the point of interaction, and require no final state interactions amongst produced partons or hadrons as modeled via scattering or fluid flow -- for a useful review see Ref.~\cite{Dusling:2015gta}.   

In the case of initial-state models, a key feature is that the particle correlations are generated within distinct color domains that have a transverse size of order $1/Q_{s}$ where $Q_{s}$ is the saturation momentum scale.  These domains extend longitudinally, thus giving rise to``ridge-like'' correlations long-range in rapidity.   It is notable that for $Q_{s} \approx 1$~GeV, the typical domain transverse size is $\approx 0.2$~fm.   Thus, even in a \pp or \pa
collision, with a saturation scale of order 1~GeV, it is possible to have a number of distinct color domains covering the interaction region.    If the various color domains have comparable field strengths, as the number of domains $N$ increases, the correlations decrease.   The reason is simply that the domains are uncorrelated in their orientation (in both coordinate space and color space) of the color fields and thus any strong angular correlation from one domain
is diluted by particles emitted from other domains with random orientations with respect to the first domain. 

In this paper, we focus in particular on the comparison of \pau and \dau collisions at RHIC because there is a clear separation of scales.   The average separation between the nucleons in the deuteron is $\langle r \rangle = 3.33$~fm which is an order-of-magnitude larger than the typical domain size.  
In the scenario where individual domains are separately resolved, there is a simple prediction that the correlation or $v_{2}$ magnitude should follow 
\begin{equation}
v_{2}(p+Au) > v_{2}(d+Au)
\label{eqn:order}
\end{equation}
since the incoherent addition of domains from the proton and the neutron in the deuteron simply increases the number of uncorrelated color fields, thereby decreasing their cumulative effect.  There appears to be consensus in the field that in the case where the individual domains of transverse size $1/Q_{s}$ are resolved, the above equation holds.   The experimental data definitively rule out this scenario~\cite{Aidala:2018mcw}.

After the submission of the full experimental data set of $v_{2}(p_T)$ and $v_{3}(p_T)$ in high-multiplicity (the highest 5\%) \pau, \dau, \hau collisions from the PHENIX collaboration~\cite{Aidala:2018mcw}, a new manuscript~\cite{Mace:2018vwq} was submitted by Mace, Skokov, Tribedy and Venugopalan (hereafter referred to as MSTV), with postdictions that appear to reconcile initial-state correlations with 
the experimental data, showing reasonable agreement with the system dependence of $v_{2}(p_T)$ (though not $v_{3}(p_T)$). Again, here we focus on the $v_{2}(p_T)$ differences between \pau and \dau for simplicity and attempt to summarize these 
surprising 
results, and then test them.
Since we are specializing to these asymmetric collisions, in what what follows ``target'' will always refer to the heavier (Au) nucleus, and ``projectile''
 will refer to the proton or deuteron. In cases where it does not cause confusion, we will refer to saturation scales in the proton that will also apply to the neutron in the deuteron. 
\section{MSTV Framework}

There are a number of calculational steps and arguments in the MSTV paper~\cite{Mace:2018vwq};  
%
here we provide only a brief summary.
The calculation is done in the dilute-dense framework, in contrast to 
previous IP-Glasma~\cite{Schenke:2012wb} calculations done in the dense-dense
framework.   Thus the proton or deuteron projectile is considered ``dilute'' and the target nucleus ``dense'' in terms of gluon occupation number.   In their calculation, MSTV consider a gluon from the target nucleus scattering from color domains
in the projectile proton or deuteron.   This implies that the nucleon is in or near a saturated gluon state where one can utilize the
weakly-coupled gluon field framework for the nucleon in a region of parton momentum fraction $x \sim 0.01$ relevant for midrapidity
hadrons produced with transverse momentum \pt $=1\mathrm{-}3$ GeV/c at RHIC.   This is in sharp contrast with the estimate in the original IP=Sat paper by Kowalski and Teaney~\cite{Kowalski:2003hm}, also at $x=0.01$, for the proton saturation scale at the
center of the proton being $Q_{s}^{2} = 0.67$~GeV$^{2}$.   Note that this is the gluon saturation value determined 
from Kowalski and Teaney's Figure~25 and the relation they provide between gluon and quark saturation scales.  
(In the text they quote a value of $1.3\ \mathrm{GeV}^2$ that appears incorrectly labeled and to be for some smaller $x$ value.)  
We highlight that $Q_{s}^{2} = 0.67$~GeV$^{2}$ is
at the center of the proton, and that integrating over a radius of 0.65 fm the average value is $\overline{Q_{s}^{2}} = 0.28$~GeV$^{2}$.
These numerical values are of interest because they are sufficiently low to 
call into question the assumption of the weak coupling 
limit for the projectile.

MSTV then state that the target gluon will interact with individually resolved color domains (in the proton) if the $k_T$ of the target gluon satisfies $k_{T} > \Qsp$, i.e., if the gluon from the target is capable of resolving domains in the projectile proton of typical (transverse) size $1/\Qsp$.   In the first draft of the MSTV paper this condition is written in
terms of $p_T$, which is also used therein for the magnitude of the transverse momentum of the final-state gluon satisfying $\mathbf{p}_T 
= \mathbf{k}_T (\mathrm{proj}) + \mathbf{k}_T (\mathrm{targ})$. 
We have benefited from private communications with the authors that have clarified that the $k_{T}$ indicated here is for the target gluon alone.

%
If the target gluon was in fact resolving individual domains in the projectile, one would have the ordering specified in Eqn.~\ref{eqn:order}, which is ruled out
by experimental data.   However, MSTV argue that if $k_{T} < \Qsp$ then the target gluon cannot resolve the individual domains in the projectile but instead interacts 
with a number  of order  $\sim \Qsp^2 /k_{T}^{2}$ of domains ``simultaneously'', also referred to by MSTV as ``coherently.''   While this may be the case it appears to be in contradiction with the requirement for the dilute-dense formalism~\cite{Dumitru:2001ux}
that $\Qsp < k_{T} < \Qst$.
In Appendix~I, we address the relative magnitude of \Qsp and \Qst and find no clear separation of scales.
This is the first of several areas of tension where we seek greater clarity in the formulation of the MSTV mechanism.

A second such concern involves the MSTV results for $v_{2}$ in \pdhau, which have an ordering $v_{2}(\dau) > v_{2}(\pau)$ up to hadron \pt $\approx$ 3~GeV and only there is there a slight hint of the ordering reverting to the pattern of the inequality in Eqn.~\ref{eqn:order}.   
Does that imply that the typical $k_T$ of a gluon from the projectile, and hence $\Qsp$, is of order $(3\  \mathrm{GeV})/2 =1.5\ \mathrm{GeV}\  \Rightarrow \Qsp^2 = 2.25\ \mathrm{GeV}^{2}$ in the proton at RHIC for $x \sim 0.01$?   As discussed below, the MSTV calculation does incorporate fluctuations in the saturation scale, but $\Qsp^2 \sim 2.25$~GeV$^{2}$ is an order of magnitude larger than the
$\overline{Q_{s}^{2}} = 0.28$~GeV$^{2}$ quoted above.
A complete understanding of the calculations presented by MSTV will require clear discussion of the numerical values of all relevant scales.   As we will see in the following discussion, it is equally important to understand the physics assumptions behind the implied coherent interactions with color fields
over the large distance scales set by the size of the deuteron.

Another key item for investigation is the assertion that with saturation scale fluctuations, 
and in the dilute-dense framework, the multiplicity $N_{ch}$ of an event, irrespective of whether it is a \pau or \dau collision, will be proportional to \QspSq.   Thus,
a 5\% highest multiplicity \dau event that has a midrapidity $dN_{ch}/d\eta \approx 18$ will have a higher saturation scale than a 5\% highest
multiplicity \pau event that has a $dN_{ch}/d\eta \approx 12$.    This is the last critical step that enables not only the ordering
of $v_{n}$ from Eqn.~\ref{eqn:order} to be negated, but in fact reversed such that $v_{2}(\dau) > v_{2}(\pau)$.   In the remainder of the paper, we test this assertion and others utilizing the \ipjazma framework.

\section{IP-Jazma Implementation}

We note at the outset that the goal of the open source \ipjazma code is not to re-implement and fully reproduce 
IP-Glasma or MSTV calculations.   In particular, in those complex numerical implementations there are too many algorithmic details and key parameters 
to be able to reproduce them in exactitude.    That project is a critical scientific
step that awaits the public release of those codes.  For these studies, 
the goal is to incorporate the identical initial physics steps, 
to gain insight on the various sources of fluctuations, 
and to test key statements in MSTV relating underlying variables.    To that end, we describe in simple language the step-by-step process in the \ipjazma calculation.   

\subsection{Monte Carlo Glauber} 
\label{Sec:MCGlauber}
\begin{figure}[hbtp]
\centering
\includegraphics[width=0.90\linewidth]{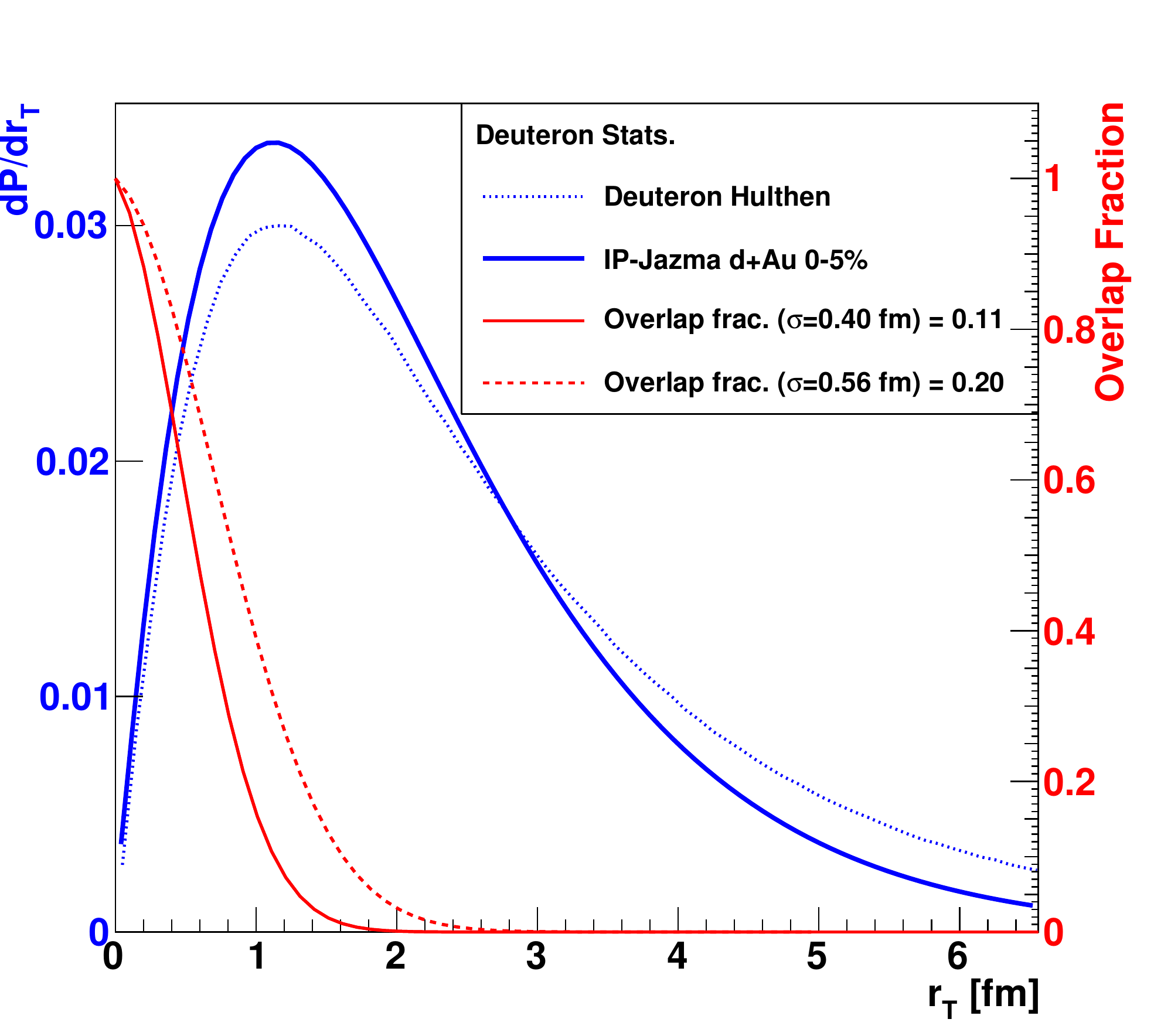}
\caption{Two-dimensional separation distance $r_T$ distribution between the proton and neutron constituents of the deuteron.   Shown are results from the deuteron Hulth\'{e}n wavefunction as well as for events selected within the 5\% highest multiplicity from the \ipjazma dilute-dense calculation.    In addition, we calculate the overlap fraction defined in Section~\ref{Sec:Overlap} as a function of $r_{T}$ between the two nucleons following two-dimensional Gaussian distributions of width $\sigma = 0.40$ and $0.56$~fm, and quote the values integrating over the \ipjazma distribution.}
\label{fig:deuteronhulthen}
\end{figure}

The first step in the \ipjazma calculation, as well as in the IP-Glasma and MSTV implementations, is to run standard Monte Carlo Glauber~\cite{Miller:2007ri} and for each collision event to output the x,y coordinates (in the plane transverse to the beam axis) of all nucleons.    In our calculation, we utilize the publicly available PHOBOS Monte Carlo Glauber code~\cite{Loizides:2014vua}.  We use the standard Woods-Saxon parameter sets in the code for the Au nucleus including the hard core repulsive parameter ($d$ = 0.4 fm), such that nucleons do not completely overlap in three-dimensional space within the nucleus.   For the deuteron the Hulth\'{e}n wavefunction is employed.   
A key insight from low-energy nuclear physics is that the deuteron is a very
loosely bound state of the proton and neutron and the average three-dimensional spatial separation between them $\langle r \rangle=3.33$~fm. In \dau collisions, the relevant length scale is the proton-neutron separation in the transverse plane $r_T$ since that defines whether both nucleons undergo inelastic collisions with the target nucleus and how far apart they strike.   
Figure~\ref{fig:deuteronhulthen} shows the distribution of $r_T$ from the Hulth\'{e}n wavefunction; the average value $\langle r_T \rangle=2.61$~fm.
The high multiplicity (0-5\% centrality) data used in the PHENIX analysis select a subset of \dau events in which both the proton and the neutron are more likely to strike near the center of the Au nucleus. The precise bias is model-dependent (for example, on the mixture of binary versus participant scaling used to model particle production), but even in the extreme case of $N_{coll}$ scaling $\langle r_T \rangle=1.66$~fm, i.e., there remains a substantial average separation between the neutron and proton.
We will further quantify these statements in the context of the \ipjazma model in the discussion to follow.

\subsection{Impact Parameter Saturation (IP-Sat)}

Next we follow the Impact Parameter Saturation (IP-Sat) model~\cite{Kowalski:2003hm} for setting the gluon saturation scale.   In this formulation, the gluon thickness function of the nucleon $T_{G}(b)$ is a function of the impact parameter, i.e. the radial distance from the center at which one probes the nucleon, and this is assumed to
have a simple Gaussian form
\begin{equation}
T_{G}(b) = \frac{1}{2 \pi B_{G}} e^{-b^{2}/(2B_{G})}
\label{Eq:ProtonProfile}
\end{equation}
where $B_{G}$ is determined via fits to electron-proton scattering data at HERA.   We highlight that there are different values of $B_{G}$ in the literature
and hence this parameter choice may be
significant.   One can then solve an implicit equation for the saturation scale squared $Q_{s}^{2}$ as a function of transverse distance from the center of the nucleon in terms of the 
strong coupling constant $\alpha_s(Q^2)$ and the gluon structure function
$g(x,Q^2)$\footnote{Note in this context $x$ is momentum fraction, not a spatial coordinate. The additional factor of $x$ in $xg(x,Q^2)$ converts $g(x,Q^2)$ to a gluon density per unit rapidity appropriate for these considerations of the saturation condition.}:
\begin{equation}
Q^{2}_{s}(x,b) =  \frac{\pi}{3R^2}  \alpha_{s} (Q^{2}_{0}+2Q^{2}_{s}) \ x g(x,Q^{2}_{0}+2Q^{2}_{s})e^{-\frac{b^2}{2B_G} } \quad .
\label{Eq:SatCond}
\end{equation}
The seemingly odd factors of $Q^{2}_{0}+2Q^{2}_{s}$ that appear here are due to the precise definition of the saturation momentum in terms of the color-dipole radius and the relation between that radius, the initial scale $Q_0$ and the color-dipole cross section found in the original IP-Sat paper~\cite{Kowalski:2003hm}; see also Appendix~I in Ref.~\cite{Schenke:2012fw}.
The value of  $R \equiv \sqrt{B_G}= 0.35$~fm is commonly used~\cite{Schenke:2012fw,Romatschke:2017ejr} which corresponds to the value $B_{G}=3.18 \pm 0.4\ \mathrm{GeV}^{-2}$ found in Ref.~\cite{Caldwell:2009ke},
where it is carefully explained that the implied small proton radius is the appropriate value for two-gluon exchange processes~\footnote{An alternative view is that this small radius characterizes the size of a constituent quark in the nucleon.}
~\footnote{Also note that this value of $B_G$ is for $x\approx 0.01$, appropriate for RHIC but not LHC energies.}.   
As detailed in Ref.~\cite{Romatschke:2017ejr}, solving Eq.~\ref{Eq:SatCond} at the $x$-scale relevant for RHIC energies, one retains to a very good approximation the Gaussian functional form -- see Figure 4.5 from that reference -- and with a slightly reduced width of $\sigma =0.32$~fm relevant for collisions at 200 GeV.  In practice, this translates into a Gaussian distribution for the squared saturation scale $Q_{s}^{2}$ with a width $\sigma$   relative to the nucleon center in the transverse plane:
\begin{equation}
Q_{s}^{2}(x,y) = Q_{s,0}^{2} \times \exp[-r_{T}^{2}/2\sigma^{2}]
\label{Eq:SatDist}
\end{equation}
where $r_{T} = \sqrt{(x-x_{i})^{2}+(y-y_{i})^{2}}$ such that $x_{i}, y_{i}$ are the center of the $i$th nucleon in the transverse plane and $Q_{s,0}^{2}$ is the squared saturation scale at the center of the nucleon.

The imperfect constraints from HERA data, together with the need to 
use values appropriate for the $x$-scale of interest, results in some ambiguity in the precise value to select for $B_G$, which then propagates into the resulting $\sigma$ for the saturation distribution (Eqn.~\ref{Eq:SatDist}).
MSTV (private communication) use the value of $B_G = 4.25\ \mathrm{GeV}^{-2}$
found in the original IP-Sat paper~\cite{Kowalski:2003hm}, resulting in
$\sigma = 0.4$~fm. It is clear that in this parameter alone there is a systematic uncertainty of order 10-15\%. There is also an unquantified systematic uncertainty in assuming the Gaussian profile of Eqn.~\ref{Eq:ProtonProfile} is valid at large distances from the center of the proton, rather than an exponential form. The study of possible exponential shapes in Ref.~\cite{Kowalski:2003hm} found the effect to be small when considering deep-inelastic scattering on the proton, but this analysis should be revisited in the current context of overlapping color domains between the proton and neutron in the diffuse deuteron configurations that still dominate the most central 0-5\% events in \dau collisions at RHIC. 
For the purposes of consistency with MSTV, we will utilize Eqn.~\ref{Eq:SatDist} with $\sigma$ = 0.40~fm throughout the remainder of this work.

We note here that the saturation scale is exactly that, a scale rather than a precise physical quantity. The convention used in Ref.~\cite{Kowalski:2003hm} to define it (and adapted by essentially all subsequent papers performing quantitative calculations) is actually framed in coordinate space: the saturation radius $r_s$ and the resulting saturation cross section is the one for which the proton represents one absorption length. The saturation momentum $Q_s$ is then defined via $Q_s^2 = 2/r_s^2$, where the factor of 2 is introduced to maintain consistency with a previous definition of a coordinate-space saturation scale by 
Golec-Biernat and W\"{u}sthoff~\cite{GolecBiernat:1999qd}. 
While there are no issues in any formulation which treats the definition of $Q_s$ consistently, it is also clear that plausible alternative definitions of $Q_s$ could differ by as much as factors of $\sqrt{2}$, so comparisons of $Q_s$ to physical momenta of real particles with similar momentum should be viewed as qualitative rather than quantitative in nature. 
Note that for the purposes of our \ipjazma calculations, this overall scale $Q_{s,0}^{2}$ is simply a normalization that will not be relevant for the
overall proportionality calculation of energy density distributions.


\subsection{$Q_{s,0}^{2}$ Fluctuations}
\label{Sec:Fluctuations}
\begin{figure}[hbtp]
\centering
\includegraphics[width=0.90\linewidth]{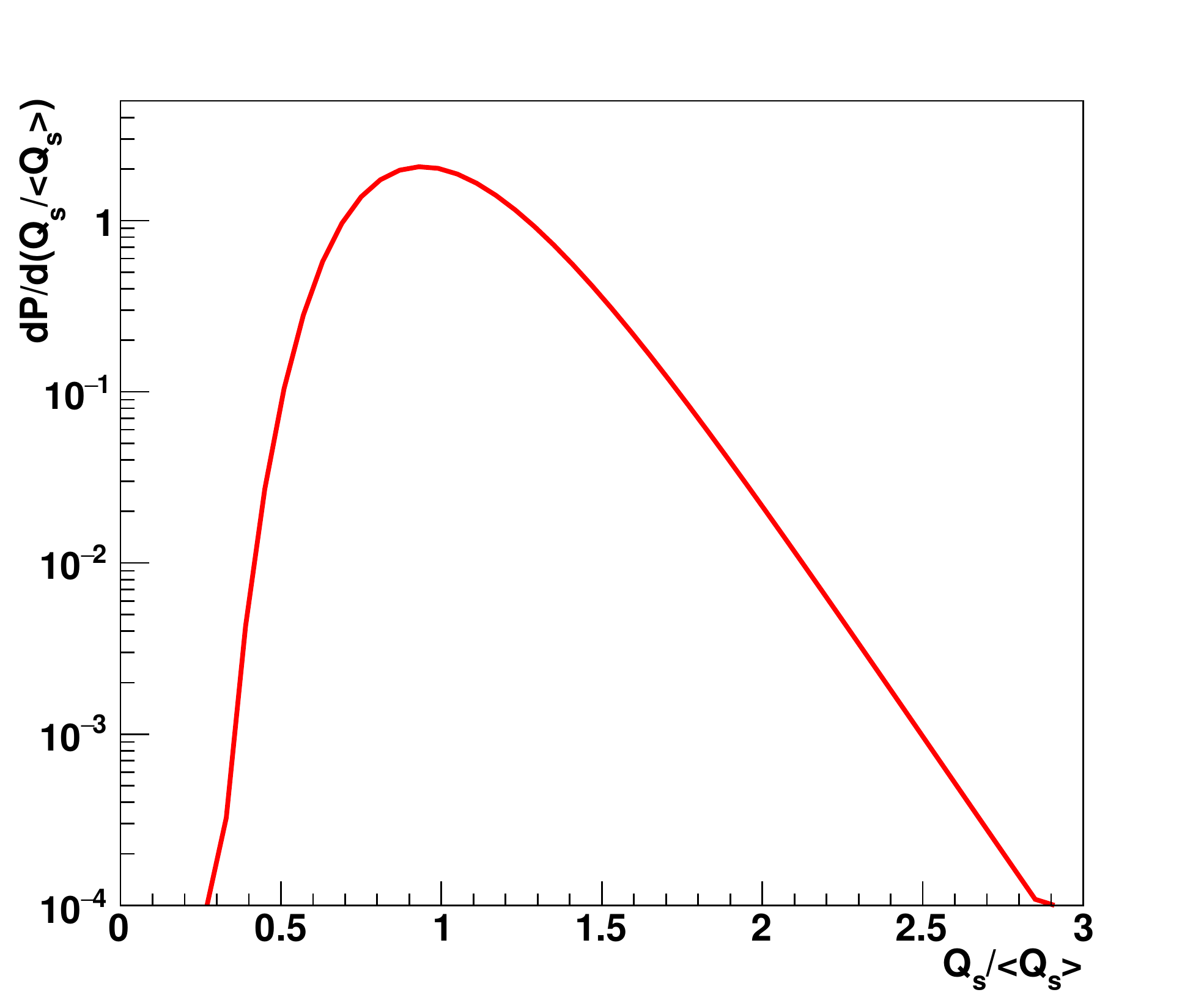}
\caption{Functional form for fluctuations in $Q_s /\langle Q_s \rangle$ with standard deviation of 0.5 on $\log[ (Q_s /\langle Q_s \rangle)^2]$.}
\label{fig:qsfluctuation}
\end{figure}

A critical component in the MSTV calculation is the inclusion of $Q_{s,0}^{2}$ fluctuations on a nucleon-by-nucleon basis.   
The theoretical basis for such fluctuations was established in 
Ref.~\cite{Iancu:2004es}. They were then calculated analytically in 
Ref.~\cite{Marquet:2006xm} and 
implemented in Ref.~\cite{McLerran:2015qxa}
in order to reproduce the $N_{ch}$ distribution in \pp collisions at the LHC in a saturation physics framework. 
Those authors argue that there may be several non-perturbative effects
that contribute to fluctuations in multiplicity, at least one of which, the event-by-event fluctuations in the saturation scale considered here, is non-perturbative and lies outside
the conventional framework of the Color Glass Condensate (CGC). 
They go on to note fluctuations in the saturation scale are critical to the original explanation of the long-range ridge in \pp collisions within the color domain picture~\cite{Dusling:2012iga,Dusling:2013qoz}. 
In order to capture these effects, 
the fluctuations are assumed to follow a log-normal distribution in the variable 
$\mathcal{Z} \equiv Q_s /\langle Q_s \rangle$
\begin{equation}
P( \mathcal{Z}  )\ d\mathcal{Z}  
=
\frac{1}{\sqrt{2\pi w^2}}\ 
\exp \left( - \frac{\log^2 ( \mathcal{Z}^2)}{2 w^2} \right)\ 
\frac{2 d\mathcal{Z}}{\mathcal{Z}}\quad .
\label{Eq:QsFlucts}
\end{equation}
with $w$ set to a value of 0.5 as used in Ref.~\cite{McLerran:2015qxa}.
The authors of Ref.~\cite{McLerran:2015qxa}
define their log-normal distribution in their Eqn.~5 in terms of 
$Q_s^2 /\langle Q_s^2 \rangle$,
but their Figure~1 is plotted as a function of 
$Q_s /\langle Q_s \rangle$. 
For the chosen value of $w=0.5$, we have
$\langle Q_s \rangle = e^{w^2/8} \sqrt{\langle Q_s^2 \rangle} = 1.03 \sqrt{\langle Q_s^2 \rangle}$, which is a negligible effect.
The resulting
distribution is shown as a function of 
$Q_s /\langle Q_s \rangle$
in Figure~\ref{fig:qsfluctuation}, and the high-side tail
from the log dependence is notable.   
In fact, the description of the high-multiplicity distribution
in \pp collisions relies on the essentially order one
fluctuation width of the distribution and the high-side tail where events with $Q_{s,0}^{2}$ up 
to 5--6 times the average value
(probabilities $6.4\mathrm{-}1.7 \times 10^{-4}$) are selected.

Although not all IP-Glasma calculations invoke such fluctuations, they are critical
in the MSTV results.   MSTV state that they fit the width of these fluctuations to minimize differences
with the STAR \dau multiplicity distribution at midrapidity~\cite{Abelev:2008ab} and obtain $w = 0.5$, the identical value used to match \pp data at 13~TeV (where the calculation is carried out in the dense-dense limit of
IP-Glasma -- see more details on this later).
It is true that the authors of Ref.~\cite{McLerran:2015qxa} comment on the (expected) slow variation of $w$ with energy, but it is striking that precisely the same value in two different formalisms is applicable at RHIC and at LHC energies.

One can  quite easily computationally incorporate (or not) such fluctuations for each nucleon in the $Q_{s,0}^{2}$  value, thus scaling up or down the entire resulting Gaussian distribution from the IP-Sat framework. Note that in doing so the width of the IP-Sat Gaussian in the transverse plane is not changed, only the overall amplitude.  The method of incorporating fluctuations in \ipjazma is similar to that Ref.~\cite{McLerran:2015qxa}, where the value of the saturation scale is fluctuated
according to Eqn.~\ref{Eq:QsFlucts}.  
However, in \ipjazma we do not perform any further sampling of local $Q_s$ and/or color charge densities on lattice points 
in the transverse plane as done in the IP-Glasma model~\cite{Schenke:2012wb}.
There is significant debate whether these fluctuations are physically well-motivated, and the improved agreement with multiplicity distributions should not be taken as evidence of such.     For example, there are many sources of fluctuations in multiplicity which are not accounted for in the Color Glass Condensate framework - consider multi-gluon jet processes - and to unambiguously attribute any missing physics to fluctuations in $Q_{s}^{2}$ that are of order 100\% with a high side tail would require additional confirmation.    It is also striking that the scale of such fluctuations identical in \dau at RHIC with $x \approx 0.01$ and in \pp at the LHC with
particle production dominated by much lower $x$.

\begin{figure*}[hbtp]
\centering
\includegraphics[width=0.97\linewidth]{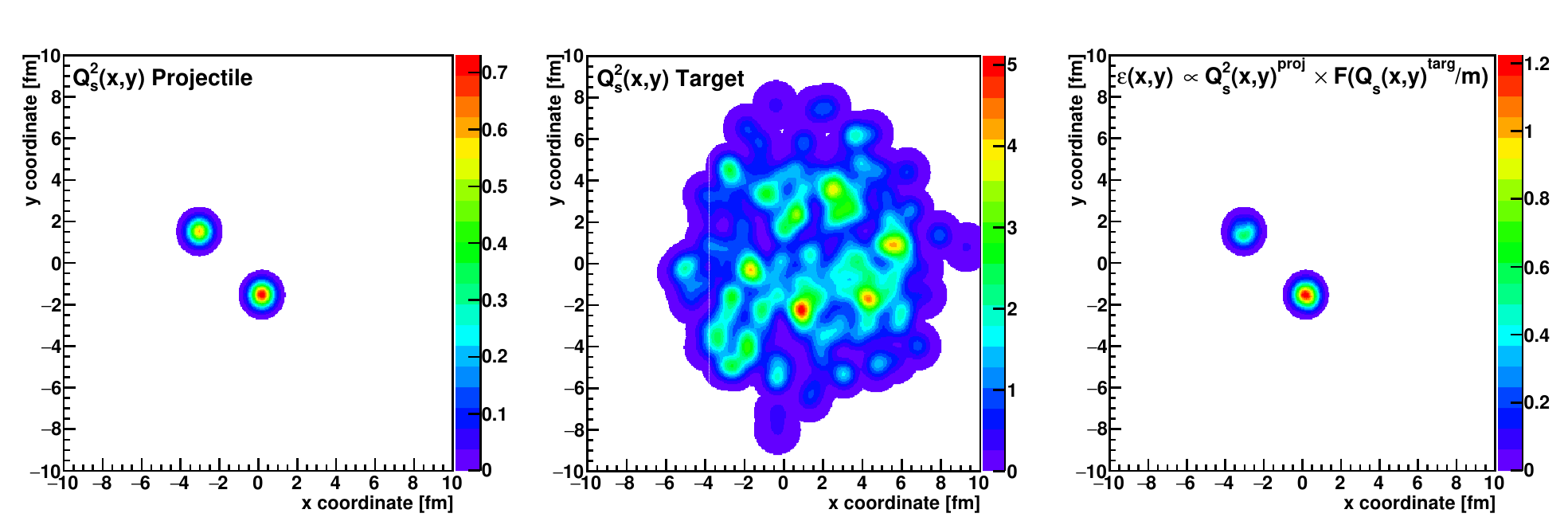}
\caption{
\ipjazma single \dau interaction event display.   Left and middle panels show the $Q_{s}^{2}(x,y)$ distribution for the projectile and target nuclei respectively.   The right panel
shows the calculated energy density from the resulting collision in arbitrary units.
  }
\label{fig:eventdisplay}
\end{figure*}

 In \ipjazma, to find the saturation scale in a collision one simply sums the $Q_{s}^{2}(x,y)$ contributions from all nucleons in the projectile to generate a two-dimensional map.
The same is done for all nucleons in the
target.   For illustration, a single \dau event at 200~GeV event is shown in Figure~\ref{fig:eventdisplay}.   The
left panel shows the two nucleons from the deuteron, each with a perfect Gaussian distribution via IP-Sat.
Note that the overall color scale (magnitude) is different for the two nucleons as they represent different
random selections from the $Q_{s,0}^{2}$ fluctuations.   The middle panel shows the summed contribution from all the nucleons in the target Au nucleus.   At this point the \ipjazma calculation should be numerically
identical to MSTV for the assumed distribution of $Q_s$ values in the transverse plane. 

\subsection{Quantifying Neutron-Proton Overlap}
\label{Sec:Overlap}
The $v_2$ ordering of Eqn.~\ref{eqn:order} relies on the domains in a \dau being resolved. This is clearly the case when the neutron and proton from the deuteron strike the Au nucleus with a 
separation between their centers exceeding the color confinement scale. Given the essentially random orientation of the deuteron in the collision, it is of interest to quantify the extent to which the color fields of the neutron and proton overlap in the ensemble of collisions that comprise the 0-5\% centrality bin in \dau collisions.
The general form of an expression for the average overlap fraction $\bar{f}_s$ for a distribution ${\cal P}(\vec{s})$ of separation in the transverse plane $\vec s$ between the proton and neutron centers is 
\begin{equation}
\bar{f}_s \equiv
\frac
{\int f_n(\vec{r}+\frac{\vec{s}}{2})\ f_p(\vec{r}-\frac{\vec{s}}{2}) \ d\vec{r} \ {\cal P}(\vec{s})\ d\vec{s} }
{\int f_n(\vec{r}                  )\ f_p(\vec{r}                  ) \ d\vec{r} \ } \quad , 
\label{Eq:Overlap}
\end{equation}
where $f_{n,(p)}(\vec{r})$ is a measure of some distribution in the neutron (proton) as 
a function of distance $\vec r$ from its center.
This is a reasonable measure of the overlap between the two nucleons; it is 1 when ${\cal P}(\vec{s}) = \delta (\vec{s})$, and will decrease with the separation $\vec s$ if $f_{n(p)}(\vec{r})$ is monotonically decreasing with $ | \vec r | $. 

To evaluate the average overlap, we use the essentially Gaussian distribution of the saturation scale in the transverse plane
\begin{equation}
f_n(\vec{r}) = f_p(\vec{r}) = \frac{Q_{s,0}^2}{2\pi \sigma^2}\ e^{-r^2/2\sigma^2}
\end{equation}
Doing this, we find that Eq.~\ref{Eq:Overlap} reduces to 
\begin{equation}
\bar{f}_s =
\int e^{-s^2/4\sigma^2}\ {\cal P}(\vec{s})\ d\vec{s} \quad .
\label{Eq:Reducedf}
\end{equation}
Again, this seems reasonable, e.g., if the two nucleon centers  are separated by $2\sigma$, we have $\bar{f}_s = 1/e$.
We have computed the numerical value of $\bar{f}_s$ as a function of $|\vec s|$
for the distribution of neutron-proton separations for 0-5\% \dau collisions plotted in Figure~\ref{fig:deuteronhulthen}
and plotted the result on the same figure (red curves). For our default value $\sigma=0.4$~fm, we find only an 11\% overlap between the neutron and proton. While extending the saturation distribution with 
$\sigma=0.56$~fm changes this value to 20\%, it is clear that the distribution of separations between the neutron and proton even in the 0-5\% most central \dau collisions is still dominated by configurations where the neutron and proton are separated by a distance exceeding the color confinement scale and therefore separately resolvable. 

\subsection{Energy Density in Dense-Dense Case}
\label{Sec:EnergyDenseDense}
In the full IP-Glasma calculation~\cite{Schenke:2012wb}, the contribution of each nucleon to the color charge squared per transverse area $g^2 \mu^2(x,\mathbf{b}_\perp)$ is assumed to be proportional to the saturation scale $Q^2(x,\mathbf{b}_\perp)$,
where $\mathbf{b}_\perp$ is the transverse projection of the impact parameter relative to each nucleon's center. 
The net color charge squared per transverse area at each lattice point in the transverse plane found by summing over the contributions of all nucleons in the projectile, and separately in the target, is used to define the rms of a Gaussian distribution for the fluctuations in local color charge. After such sampling, the resulting charge distributions are used to calculate the electric and magnetic color fields by solving the Classical Yang-Mills (CYM) equations. The local value of the energy density is then computed from the gluonic fields. 
It is the  lattice site fluctuations in local color charge density which give the IP-Glasma event displays of energy-density their fine-scale spiky visual features.  As noted in Ref.~\cite{Romatschke:2017ejr}, the initial spatial scale for the color 
fluctuations is that of the lattice spacing and therefore not entirely physical.

The authors of Ref.~\cite{Romatschke:2017ejr} follow the full IP-Glasma formalism
and find that, after averaging over these color fluctuations, one obtains a remarkably simple answer.  
The resulting energy density at proper time $\tau=0$ for each lattice site is given by:
\begin{equation}
\varepsilon(x,y) \propto g^{2} Q_{s}^{2}(x,y)_{proj} \times Q_{s}^{2}(x,y)_{targ}
\label{eqn:densedense}
\end{equation}
where $g$ is the strong coupling. 
(Important discussions on the time evolution away from $\tau=0$ and the dependence on the lattice spacing may be found in Ref.~\cite{Lappi:2006hq}.)
We note that in some of the saturation physics literature, factors
of $g$ are absorbed into the definition of the saturation scale.

This result is essentially number of collisions $N_{coll}$ scaling for the energy density, i.e. it is the product of the projectile and target thickness functions -- only the modest non-linearities in Eqn.~\ref{Eq:SatCond} prevent this from being strictly true.   Note that $N_{coll}$ scaling in traditional Monte Carlo Glauber calculations treat all nucleon-nucleon binary collisions equally; as a result scaling by $N_{coll}$ does not match the expected energy density distribution.   However, in this case,
the thickness functions include a version of an impact parameter dependence for each nucleon-nucleon interaction, i.e. more peripheral N-N interactions have a smaller overlap of their IP-Sat Gaussian distributions.    In that sense, this physics scaling is very similar to Monte Carlo Glauber with constituent quarks that essentially give an impact parameter dependence to nucleon-nucleon interactions~\cite{Loizides:2016djv}.   
In the publicly available {\sc{TRENTO}} model~\cite{Moreland:2014oya}, the authors populate arbitrary Gaussian distributions for projectile and target nucleons and find, in the so-called ``$p$=0 geometric'' mode, that the resulting energy density is proportional to $\sqrt{T_{A} \times T_{B}}$, where $T_{A,B}$ are the nuclear thickness functions.    The square-root is arbitrary and as formulated does not represent $N_{coll}$ scaling, but again this approach does incorporate a variant of an impact parameter
dependence for N-N interactions.  Thus, for certain parameter selections {\sc{TRENTO}} $p=0$ approximately reproduces the IP-Glasma \nucnuc eccentricities $\varepsilon_{2}, \varepsilon_{3}$ -- see Figure 3 from Ref.~\cite{Moreland:2014oya}.

\begin{figure*}[hbtp]
\centering
\includegraphics[width=0.97\linewidth]{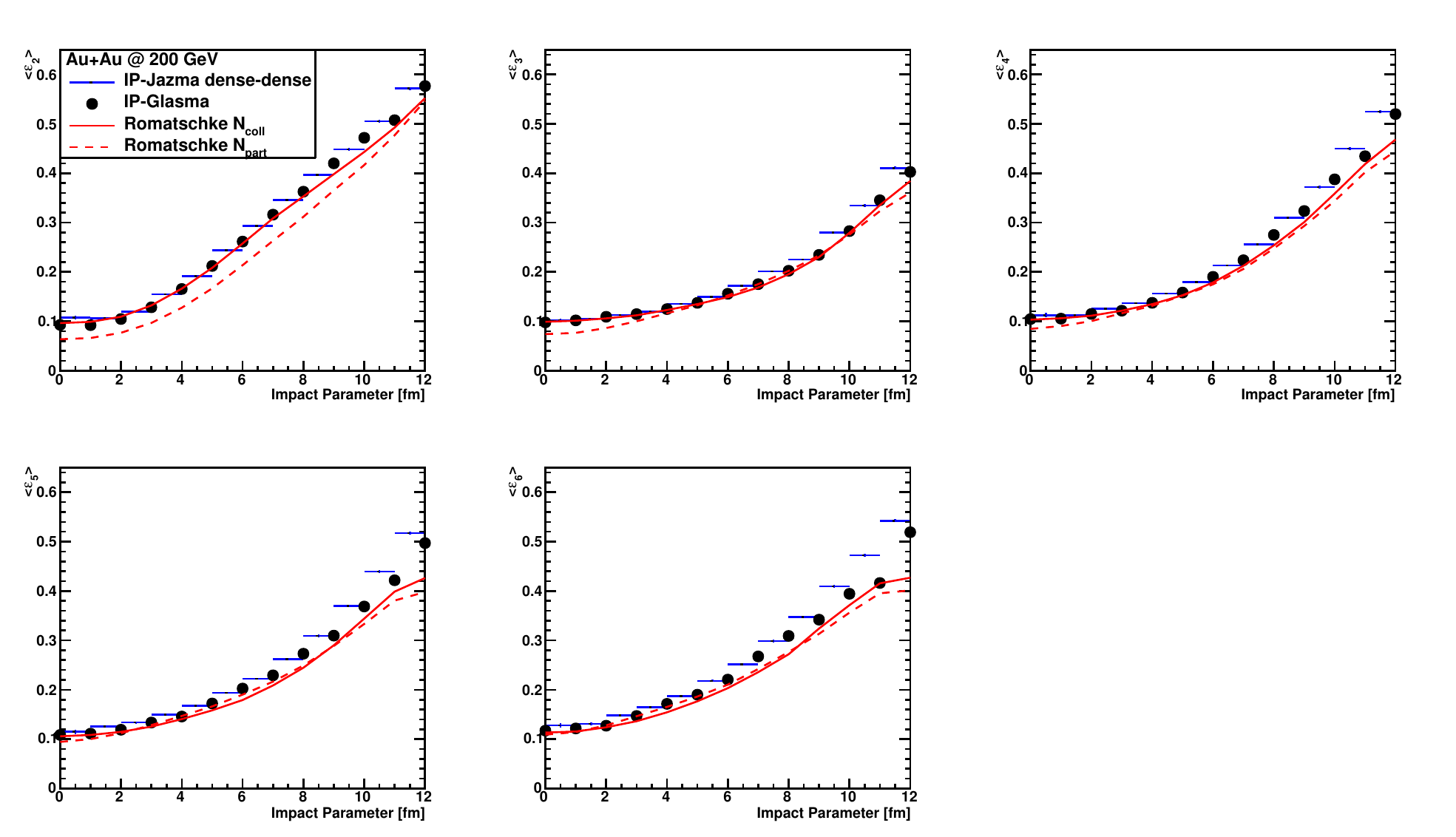}
\caption{
Calculations of eccentricity moments $\varepsilon_{2}\mathrm{-}\varepsilon_{6}$ for Au+Au collisions at 200 GeV.
Shown are results from the full IP-Glasma calculation~\cite{Schenke:2012fw} 
compared with calculations from the \ipjazma 
code.    Also shown are calculations from Ref.~\cite{Romatschke:2017ejr} in the $N_{coll}$ and $N_{part}$ cases, noting that the $N_{coll}$ case is the same algorithm as in \ipjazma modulo issues such as the 
$r_{max}$ cutoff.
}
\label{fig:auauecc}
\end{figure*}

The result in Eqn.~\ref{eqn:densedense} is obtained from the full \hbox{IP-Glasma} gluon field calculation followed by averaging over the fluctuations induced by the Gaussian sampling of color charge on each lattice site. Typical lattice grids used in IP-Glasma calculations are of  order $0.025 \times 0.025$ fm$^{2}$.  One can ask quantitatively when and if these fluctuations are a significant or insignificant contributor to various physics observables -- a question that is not asked often enough.   For example, Figure~4.7 from Ref.~\cite{Romatschke:2017ejr} shows for \auau collisions at 200 GeV the eccentricities 
$\varepsilon_{n}$ for $n=2\mathrm{-}6$ obtained after averaging over the lattice-sized color fluctuations match almost perfectly with the full IP-Glasma calculations~\cite{Schenke:2012fw}.   This is
particularly notable for impact parameter $b=0$ where fluctuations dominate.    We reproduce those results in Figure~\ref{fig:auauecc} and compare to \ipjazma results in the dense-dense limit using the simple result in Eqn.~\ref{eqn:densedense}. Overall, there is very good agreement between the IP-Glasma and \ipjazma results for the various eccentricities.   For more peripheral collisions $b > 8$~fm, there are small deviations for the higher moments $n \ge 4$ which could be related to these additional lattice site fluctuations or parameter choices such as the assumed maximal extent of the IP-Sat Gaussian distribution, as discussed below.

There are a few additional items to mention in this regard.   Functionally in the algorithm there is a choice for how far to extend the IP-Sat Gaussian, referred to as $r_{max}$.     Calculations can be
sensitive to $r_{max}$ and often extend this to the edge of the entire lattice grid.   This choice can influence the eccentricities which could also help explain the differences
in more peripheral collisions mentioned above.   The IP-Sat assumption of a Gaussian form and parameter
setting from HERA data is very unlikely to have any constraint on the tail of the distribution for distances from the center of the nucleon exceeding $(2\mathrm{-}3)\sigma\sim (0.8\mathrm{-}1.2)\ \mathrm{fm}$ and thus any observable sensitive to choices in $r_{max}$ beyond this must be viewed as
systematic uncertainties.    In the \ipjazma case, we set $r_{max} = 3.0 \times \sigma$ in all results
shown here.    
In the comparison for \auau eccentricities as a function of impact parameter mentioned above , there is also the question of what defines the limit of an inelastic collision.    If one extends the IP-Sat Gaussian out further, one effectively has a larger inelastic cross section.    Various schemes for matching the experimental total \nucnuc inelastic cross section are discussed in Ref.~\cite{McDonald:2016vlt}.
One last item is that in some papers, the factor $g^{2}$ in Eqn.~\ref{eqn:densedense} is allowed to run with $Q^2$
and is evaluated at the maximum value on the lattice site between $Q_{s}$(proj) and $Q_{s}$(target).   This
is not standard across IP-Glasma papers, and the $g^{2}$ is treated as a constant in the \ipjazma
calculations shown in this paper.
 
Before proceeding to a discussion of the dilute-dense limit used in MSTV, we note that there have been several calculations for small systems using the dense-dense limit in the IP-Glasma framework, including setting initial conditions for \pdha collisions~\cite{Bzdak:2013zma,Schenke:2014gaa} and calculating small system multiplicity and momentum distributions~\cite{Schenke:2013dpa,Schenke:2018hbz}.

\subsection{Gluon Density in Dilute-Dense Case}
The calculation of MSTV is performed in the dilute-dense limit of saturation physics. 
The authors use the same procedure of Monte Carlo 
Glauber and IP-Sat~\cite{Kowalski:2003hm} deployed in the IP-Glasma framework~\cite{Schenke:2012wb}, including the treatment of fluctuations.  
However, for the subsequent evolution of the gluon field 
they employ the dilute-dense formalism. appropriate for 
small systems incident on heavy targets.
While the dilute-dense limit was initially developed for \ppb collisions~Refs.~\cite{Blaizot:2004wu,Blaizot:2004wv},
it is applicable whenever one system (the
projectile) has a saturation scale significantly lower than that of the other system (the target), i.e. 
$Q_{s}$(proj)$ < Q_{s}$(targ).
Care must be taken to ensure correct treatment of the odd angular harmonics
necessary for the generation of $v_3(p_T)$~\cite{McLerran:2016snu}.
 In general, the dilute-dense formalism
is considered valid when the $Q_{s}$(proj)$ < k_{T} < Q_{s}$(targ).
As noted at the end of the previous section, while this may seem quite natural for the treatment of \pau and \dau collisions at RHIC, reasonable results have also been obtained working in the dense-dense limit for these systems.    

\begin{figure}[hbtp]
\centering
\includegraphics[width=0.9\linewidth]{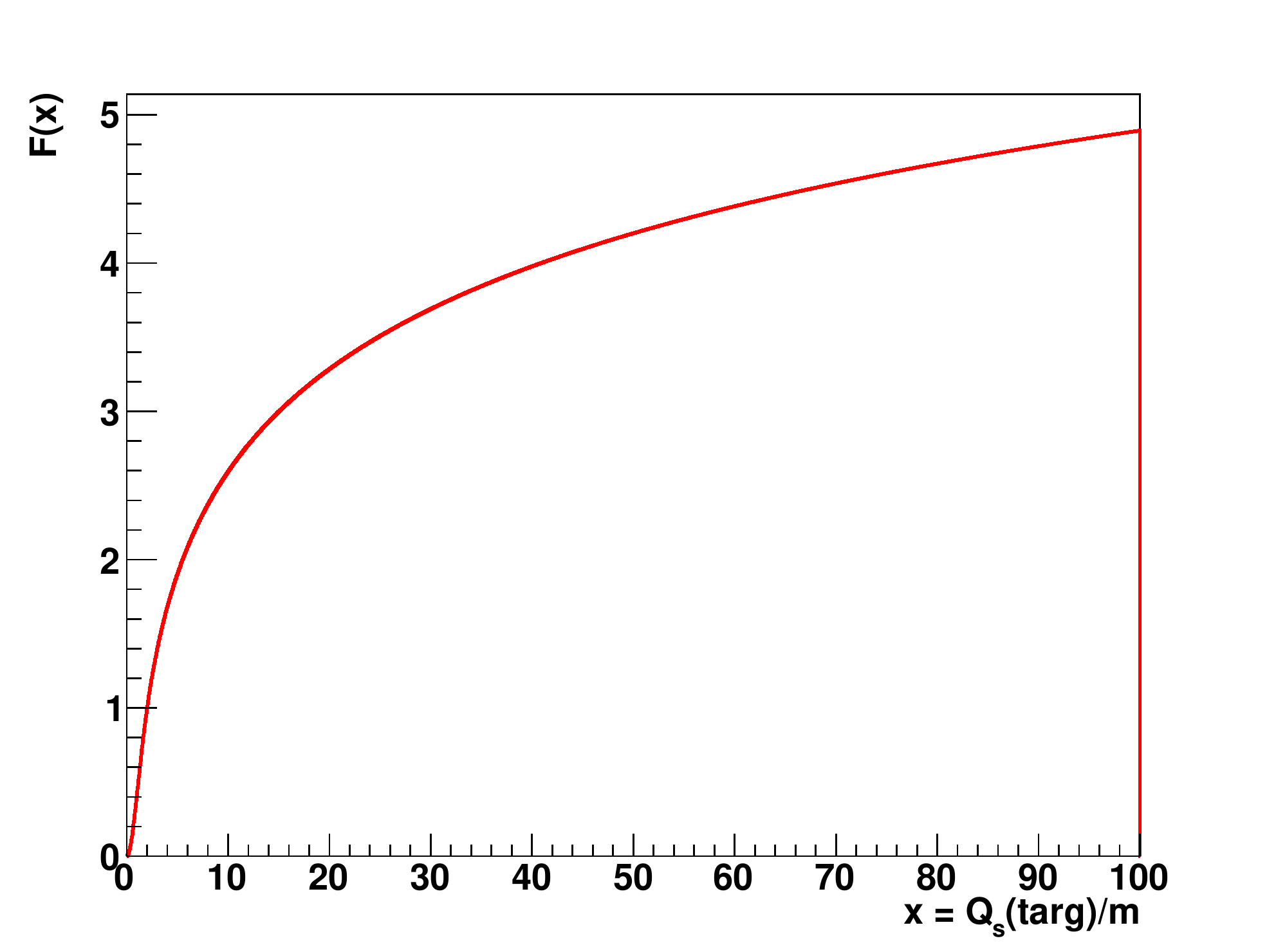}
\caption{Functional form $F(x)$ where $x=Q_{s}(targ)/m$.}
\label{fig:cyrille}
\end{figure}

In the dilute-dense formalism, after averaging over color fluctuations,  the local gluon density is given by
\begin{equation}
N_g(x,y) \propto g^{2} Q_{s}^{2}(x,y)_{proj} \times F(Q_{s}(x,y)_{targ}/m)
\label{eqn:dilutedense}
\end{equation}
where $m$ is the infrared cutoff applied in the calculation.  
The function $F(u)$ is taken as
\begin{equation}
F(u) = \int_{0}^{u} dy\  [1-e^{-y^{2}}]/y
\label{eqn:cyrille}
\end{equation}
where $u = Q_{s}$(targ)$/m$~\footnote{Cyrille Marquet, private communication.}.  
As shown in Figure~\ref{fig:cyrille}, 
the function $F(u)$ is clearly logarithmic in $(Q_{s}/m)$ at large values of $Q_{s}/m$.   

We take the same numerical value $m = 0.3$~GeV as used in MSTV, which was selected in order to best match the \dau multiplicity distribution.    MSTV cites an earlier IP-Glasma paper relating to systematic uncertainties from this parameter variation, though in that paper they only vary $m$ from 0.1 -- 0.2~GeV.
We note that in  \ipjazma we find very little sensitivity to this parameter.
Figure~\ref{fig:eventdisplay} (right panel) shows a single \dau event and the energy density in arbitrary units calculated using this dilute-dense formulation.

\subsection{Gluons and Energy Density}

For calculating initial spatial eccentricities, the above formalism is complete within \ipjazma.   
For matching experimentally measured charged hadron multiplicity distributions, \ipjazma faces the same issues confronting any theoretical model.
The experiments do not measure all neutral hadrons, there are 
experimental acceptance and efficiency effects including a low-\pt cutoff, and there are mapping issues
from gluons to hadrons and associated fluctuations.    
Renormalizing the distributions relative to the 
mean quantity, e.g. $N_{g}/\langle N_{g} \rangle$ can ameliorate some of these effects, but does not eliminate issues in the shape of the distribution.
For example, an experiment measuring hadrons over two units of rapidity will on average measure twice the particles relative to an experiment measuring over one unit (assuming one is on the rapidity plateau).   However, rescaling the distribution by a factor of two will not bring the shapes into agreement as there will be a wider distribution in the smaller acceptance case.

That said, for comparing with the MSTV results, within \ipjazma, we simply assume that the number of gluons is linearly proportional to the the energy density, and therefore will take  Eqn.~\ref{eqn:densedense} and Egn.~\ref{eqn:dilutedense} (after summing over lattice sites)
as proportional to the number of charged hadrons in the applicable dense-dense and dilute-dense limits respectively.
 In this initial treatment we defer questions of energy density versus entropy density, given our reasonable description of the \dau multiplicity distribution (Figure~\ref{fig:dau_dilutedense_yesfluc}), and our goal of understanding the role of various sources of fluctuations in the MSTV calculation.  
While there are multiple places where fluctuations come into these calculations,   
it is our intent here to test
the contributions of fluctuations just from Monte Carlo Glauber and IP-Sat $Q_{s,0}^{2}$ fluctuations in order to understand the importance of other sources of fluctuations within MSTV.


\section{\ipjazma Results}
\label{Sec:IPJazmaResults}
We begin with the simplest results from \ipjazma for light systems and then systematically explore their implications.    First consider \pau collisions at RHIC treating the system in the dilute-dense limit, with no $Q_{s,0}^{2}$ fluctuations.
In all cases the width of the IP-Sat Gaussian $\sigma = 0.40$~fm, the $r_{max} = 3.0\times \sigma$ within IP-Sat, the infrared regulator $m=0.3$~GeV and we assume a constant value for $g^{2}$ as it appears in Eqn.~\ref{eqn:dilutedense}.   
Note that the exact numerical
value of the average $Q_{s,0}^{2}$ does not enter since we will compute the number of gluon distribution
relative to the average number of gluons ($N_{g}/\langle N_{g} \rangle$) as done in MSTV.   
For distributions showing the dependence of $Q_{s,0}^{2}$ with various parameters 
we have chosen $Q_{s,0}^{2} = 0.67$~GeV$^{2}$ as per previous discussion of this value taken from Ref.~\cite{Kowalski:2003hm}.   

\begin{figure*}[hbtp]
\centering
\includegraphics[width=0.45\linewidth]{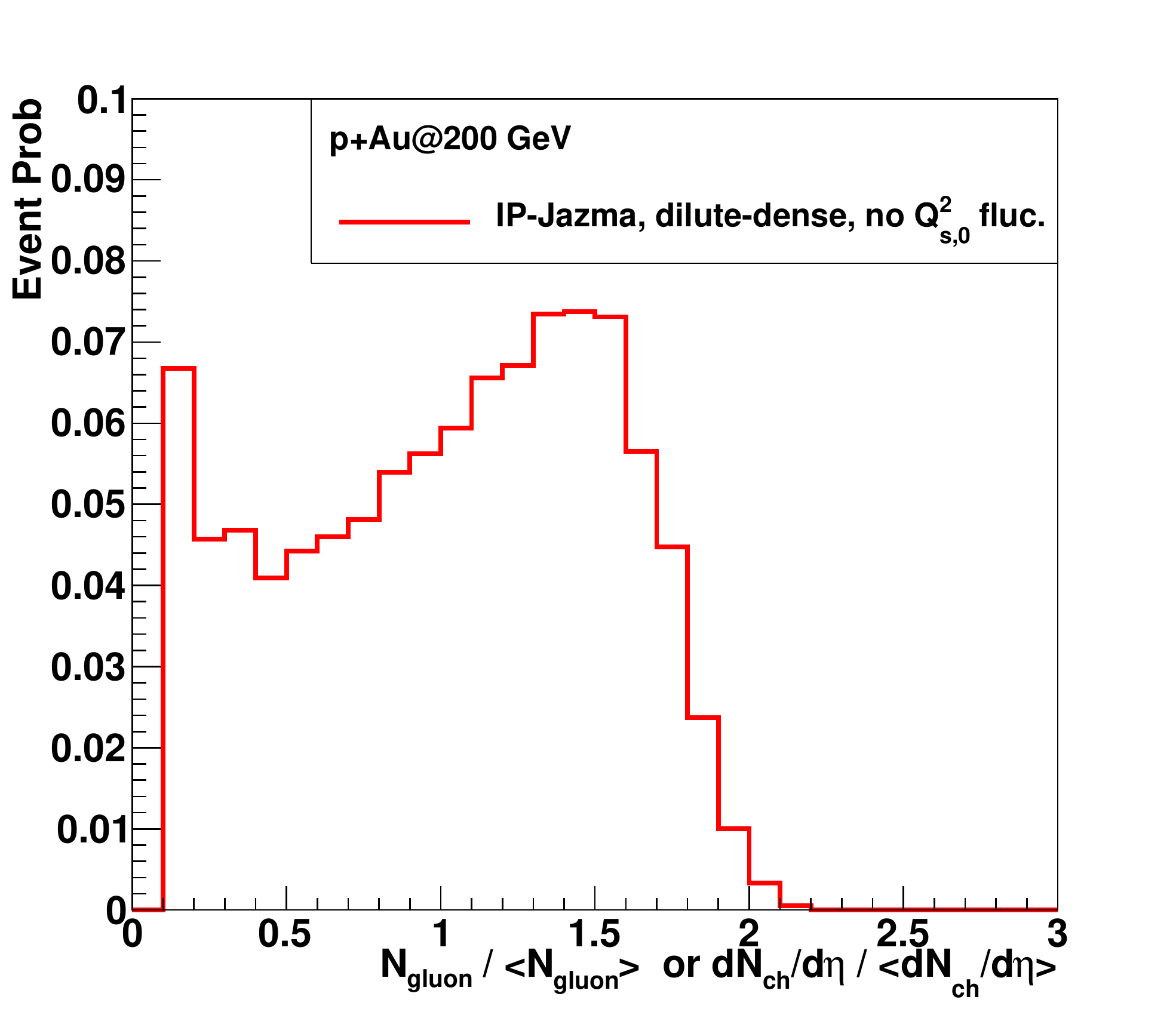}
\includegraphics[width=0.45\linewidth]{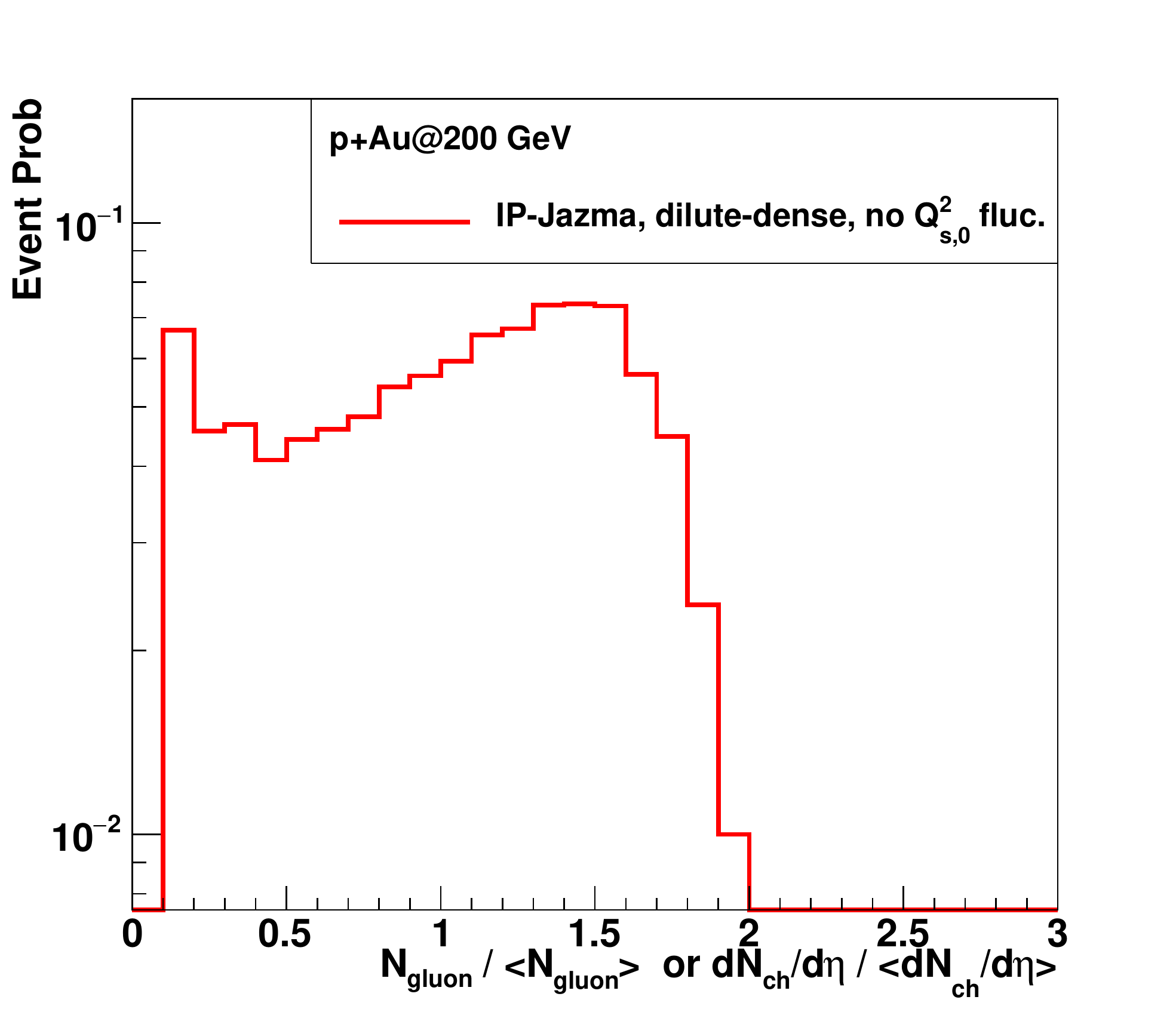}
\caption{\ipjazma result for the distribution of $N_{g}/\langle N_{g} \rangle$ in the dilute-dense case and with no 
$Q_{s,0}^{2}$ fluctuations.  The left (right) panel has the y-axis on a linear (log) scale.}
\label{fig:pau_dilutedense_nofluc}
\end{figure*}

The resulting distribution for $N_{g}/\langle N_{g} \rangle$ in \pau collisions is shown
in Figure~\ref{fig:pau_dilutedense_nofluc}.
The distribution has a peak for very low gluon number, in cases where the proton strikes the edge of the
nucleus.   The distribution then has a somewhat stronger peak near the maximum value, relative to the mean, corresponding to
those cases where the proton hits the ``thick-enough'' part of the target nucleus to ``free'' all the gluons in the proton.   For such configurations, the number of gluons is only logarithmically dependent on the target
thickness and one cannot generate any more multiplicity.   

\begin{figure*}[hbtp]
\centering
\includegraphics[width=0.9\linewidth]{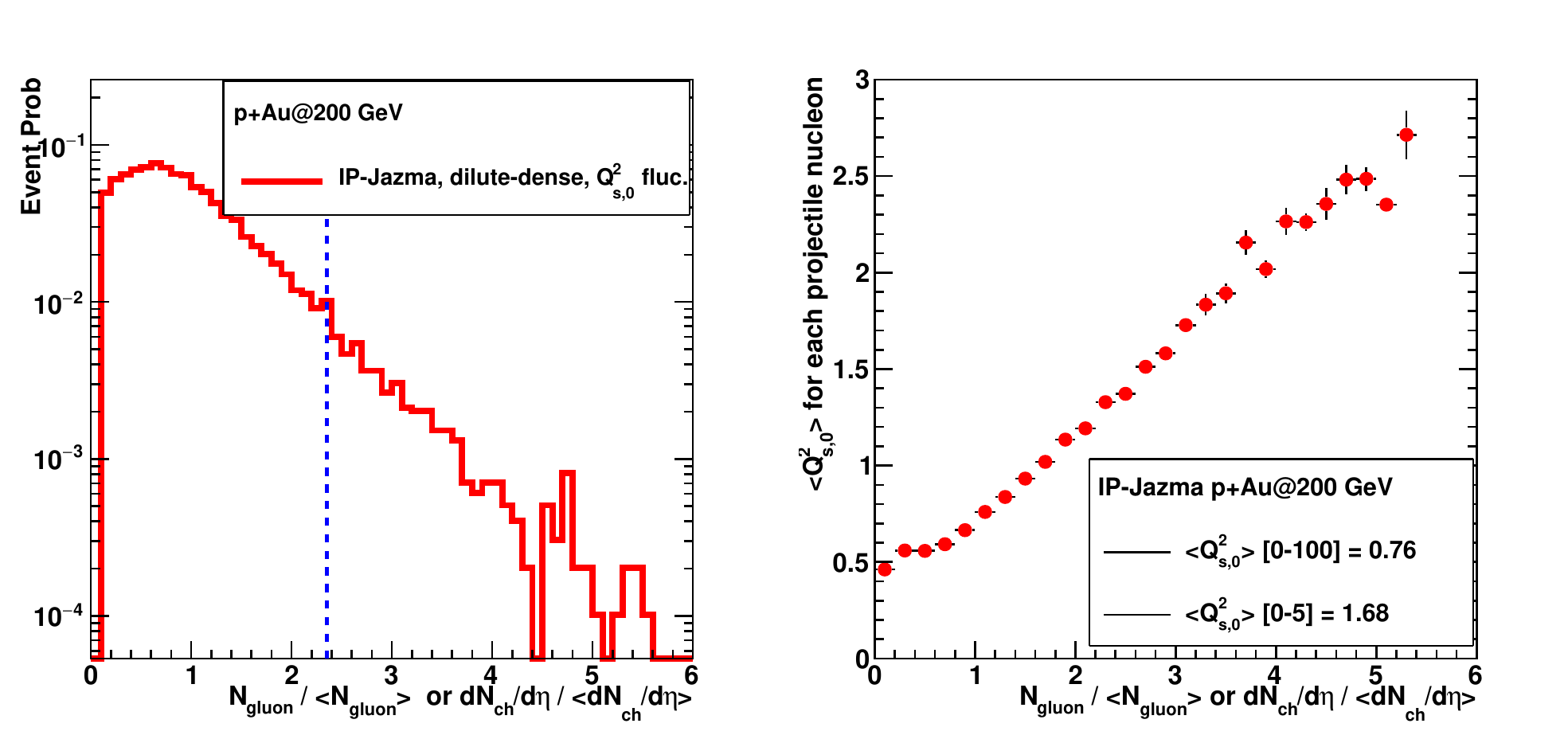}
\caption{\ipjazma results in \pau collisions for the distribution of $N_{g}/\langle N_{g} \rangle$ in the dilute-dense case and with the 
inclusion of $Q_{s,0}^{2}$ fluctuations.  The blue dashed line indicates the cutoff for the
5\% highest multiplicity events.  The right panel shows the average projectile proton 
$Q_{s,0}^{2}$ as a function of event selected $N_{g}/\langle N_{g} \rangle$.}
\label{fig:pau_dilutedense_yesfluc}
\end{figure*}

Now we perform the identical calculation but with the inclusion of $Q_{s,0}^{2}$ fluctuations for all nucleons --
in both the projectile proton and the target nucleons.   
The resulting gluon distribution is shown in 
Figure~\ref{fig:pau_dilutedense_yesfluc} (left panel).    One immediately sees that the shape of the distribution is qualitatively different, being effectively dominated by the choice of magnitude and shape of the $Q_{s,0}^{2}$ fluctuations.   The blue
dashed line indicates the selection on the highest 5\% multiplicity events.    

Since this is a Monte Carlo calculation,  
we can calculate the average value for $Q_{s,0}^{2}$ in the proton
for all events falling into a particular gluon multiplicity selection.   These values are shown in 
Figure~\ref{fig:pau_dilutedense_yesfluc} (right panel).   Again, the mean value is arbitrary at this point;
the key take-away message is that the gluon multiplicity essentially depends linearly on the proton
$Q_{s,0}^{2}$ value.    The only deviation is at low multiplicity when the proton hits the edge of the
nucleus.   This is completely consistent with the statement in MSTV that in the dilute-dense limit there is
this simple proportionality $N_{gluon} \propto Q_{s,0}^{2}$.    In selecting the highest 5\% multiplicity
events, we find in \ipjazma that in \pau collisions at RHIC the average $Q_{s,0}^{2}$ is higher by a factor of 1.68/0.76 = 2.2 than the
average.

\begin{figure*}[hbtp]
\centering
\includegraphics[width=0.45\linewidth]{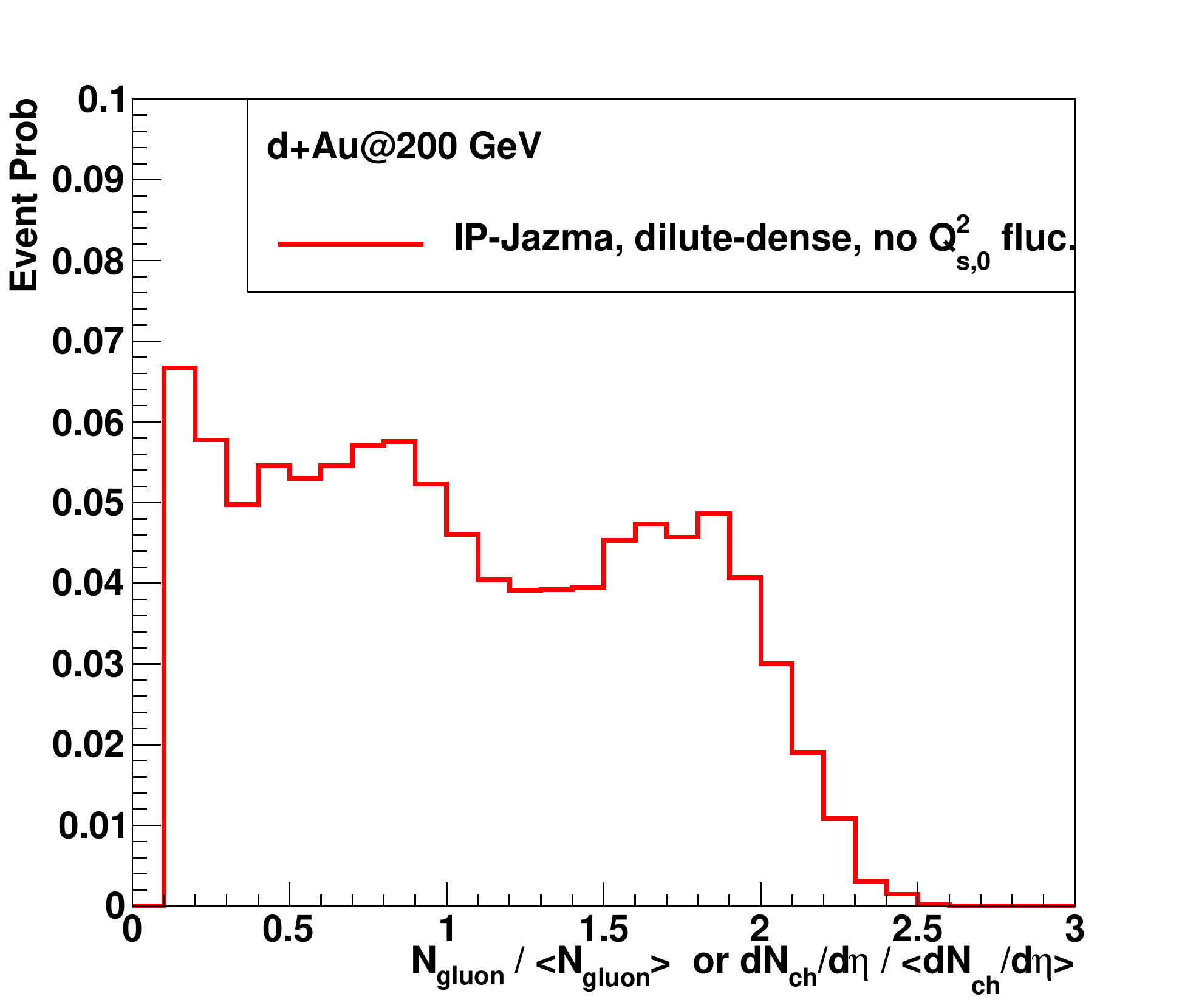}
\includegraphics[width=0.45\linewidth]{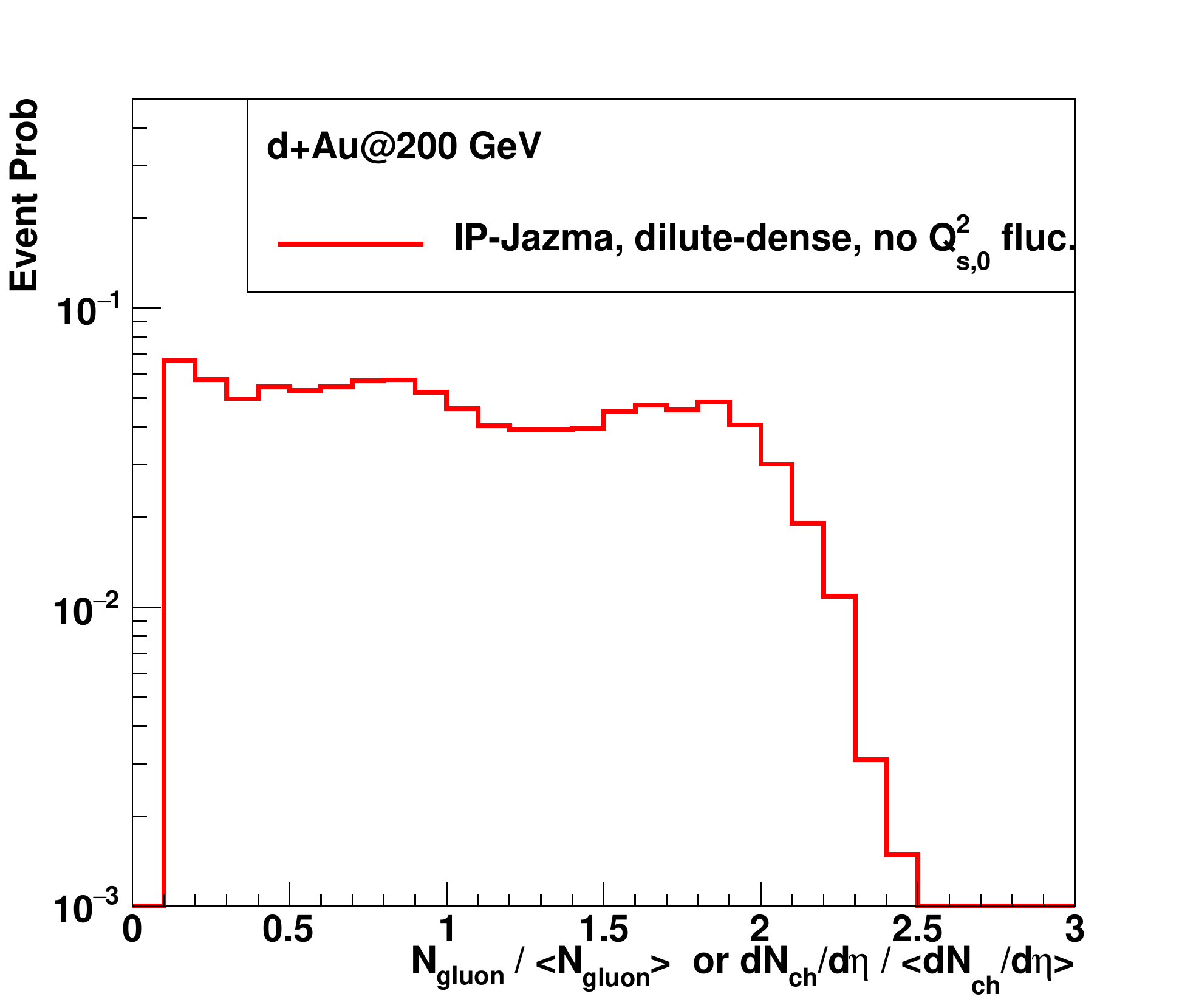}
\caption{\ipjazma \dau minimum bias results for the distribution of $N_{g}/\langle N_{g} \rangle$ in the dilute-dense case and with no $Q_{s,0}^{2}$ fluctuations.  The left (right) panel has the y-axis on a linear (log) scale.}
\label{fig:dau_dilutedense_nofluc}
\end{figure*}

We now move to the \dau case and show in Figure~\ref{fig:dau_dilutedense_nofluc} the distribution of $N_{gluon}/\langle N_{gluon} \rangle$ in the dilute-dense case and without $Q_{s,0}^{2}$ fluctuations.   In this case, the
distribution has two peaks away from zero.   The peak around $N_{g}/\langle N_{g} \rangle \approx 0.8$ corresponds to when only one nucleon from the deuteron hits the target nucleus and in a thick enough region to
fully free the projectile (single nucleon) gluons.
Due to the large size of the deuteron, the relative size of this first peak is substantial. 
The other peak around $N_{g}/ \langle N_{g} \rangle \approx 1.7$ is produced when both nucleons from the deuteron hit the thick region of the target nucleus.

\begin{figure*}[hbtp]
\centering
\includegraphics[width=0.9\linewidth]{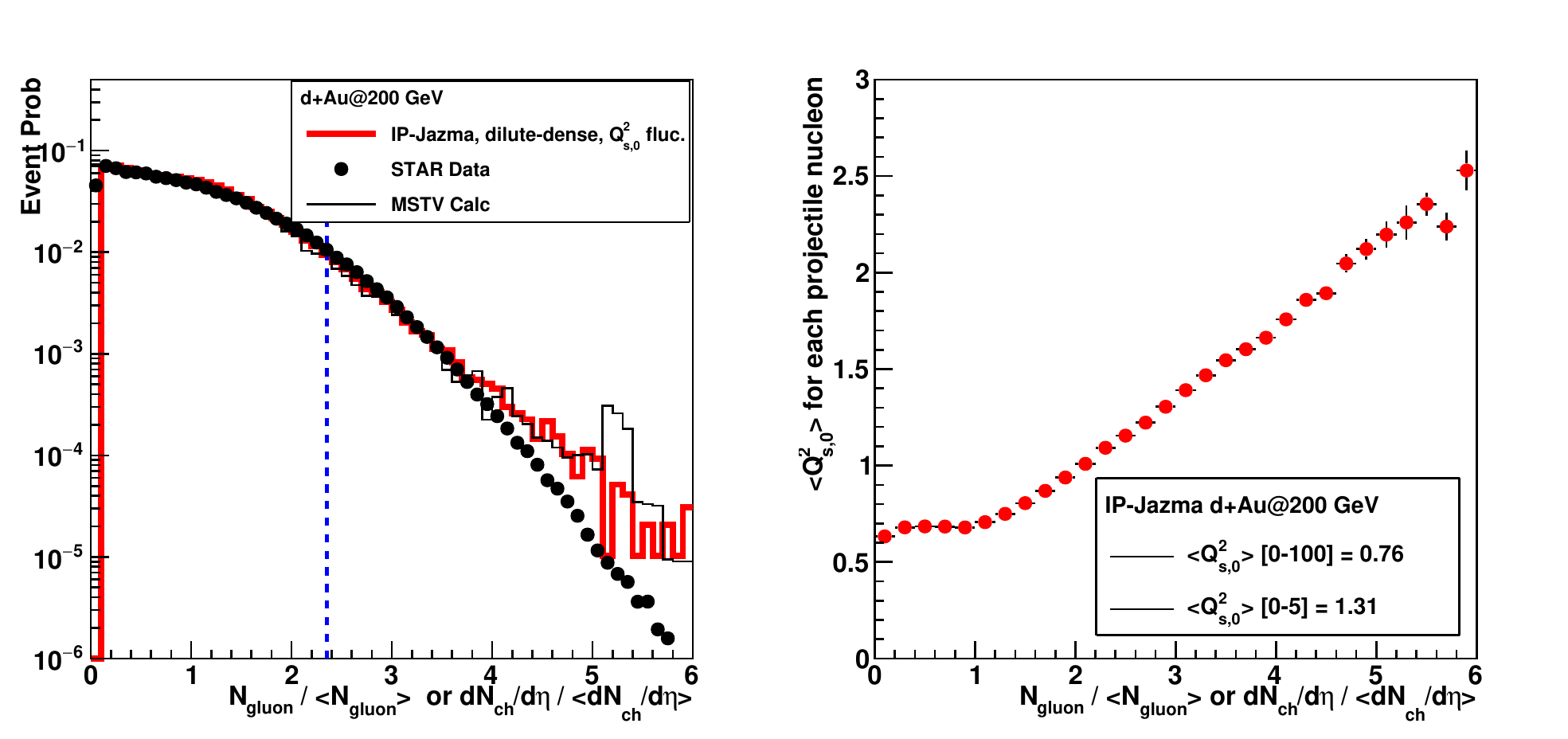}
\caption{\ipjazma results in \dau collisions for the distribution of $N_{g}/\langle N_{g} \rangle$ in the dilute-dense case and with the 
inclusion of $Q_{s,0}^{2}$ fluctuations.  The blue dashed line indicates the cutoff for the
5\% highest multiplicity events.  The right panel shows the average projectile proton 
$Q_{s,0}^{2}$ as a function of event selected $N_{g}/\langle N_{g} \rangle$.}
\label{fig:dau_dilutedense_yesfluc}
\end{figure*}

Next we calculate the gluon distribution for \dau collisions in the dilute-dense case and with the
prescribed $Q_{s,0}^{2}$ fluctuations, shown in Figure~\ref{fig:dau_dilutedense_yesfluc}.
The key observation is that the \ipjazma results agree almost perfectly with the MSTV calculation.  
This \ipjazma result in itself is quite remarkable and
indicates that by far the dominant source of fluctuations come from Monte Carlo Glauber in combination with 
$Q_{s,0}^{2}$ fluctuations which are non-perturbative and lie outside the CGC framework~\cite{McLerran:2015qxa}.
This is in sharp contradistinction to other sources related to color fluctuations
and any ``remarkable" derivation of negative binomial fluctuations in the Color Glass Condensate framework~\cite{Gelis:2009wh}.
(See also Appendix~II in this regard.)
Note that in the \ipjazma result there are absolutely zero free parameters in the sense that each numerical value, where applicable, 
exactly matches those used by MSTV.

Also shown in Figure~\ref{fig:dau_dilutedense_yesfluc} are charged hadron data within $|\eta| < 1.0$ from the STAR collaboration~\cite{Abelev:2008ab}.   The MSTV calculation was matched to this distribution with three parameters, the infrared cutoff $m = 0.3$~GeV, the width $w=0.5$ of the $Q_{s,0}^{2}$ fluctuations in $\log[ ( Q_{s,0} / \langle Q_{s,0} \rangle\ )^2 ] )$, and the scale factor relating the saturation scale and the color charge 
density~\cite{Mace:2018vwq}.    Both  \ipjazma and MSTV reproduce the data reasonably, with the only
discrepancy being the significant over-prediction in both for the 0.5\% highest multiplicity events.    
We have verified in \ipjazma that this over-prediction is directly related to the width $w$ in Eqn.~\ref{Eq:QsFlucts}; changing $w=0.5 \rightarrow 0.45$ provides a good description of the very high multiplicity tail, at the expense of slightly worsened agreement with the data in the region $N_g / \langle N_g \rangle \sim 1\mathrm{-}4$. 

In the right panel of Figure~\ref{fig:dau_dilutedense_yesfluc} we show the average $Q_{s,0}^{2}$ for
the neutron and proton from the projectile deuteron as a function of event multiplicity category.  As in
the \pau case, there is a linear relationship between $N_{gluon}$ and $Q_{s,0}^{2}$ once both projectile
nucleons move inside the edge of the target nucleus.    The 5\% highest multiplicity \dau events have an enhancement in $Q_{s,0}^{2}$ by 1.31/0.76 = 1.7, which is lower than the enhancement ratio of 2.2 found in the 5\% highest multiplicity \pau events.   This is to be expected from
basic probability arguments 
under the assumption that the neutron and proton in the deuteron fluctuate separately.
In these figures, we have followed MSTV in presenting the multiplicity results after scaling  $ \langle N_{gluon} \rangle $, but of course we know the
numerical value for the average value.   
We find in \ipjazma that the 0-5\% \dau events have a multiplicity that is 1.5 times
higher than the 0-5\% \pau events.   This is very consistent with the ratio of PHENIX experimental
measurements of $dN_{ch}/d\eta$ at midrapidity between the 5\% most-central \dau and \pau events.

\section{Discussion and Analysis}
\label{Sec:DandA}

\begin{figure*}[hbtp]
\centering
\includegraphics[width=0.9\linewidth]{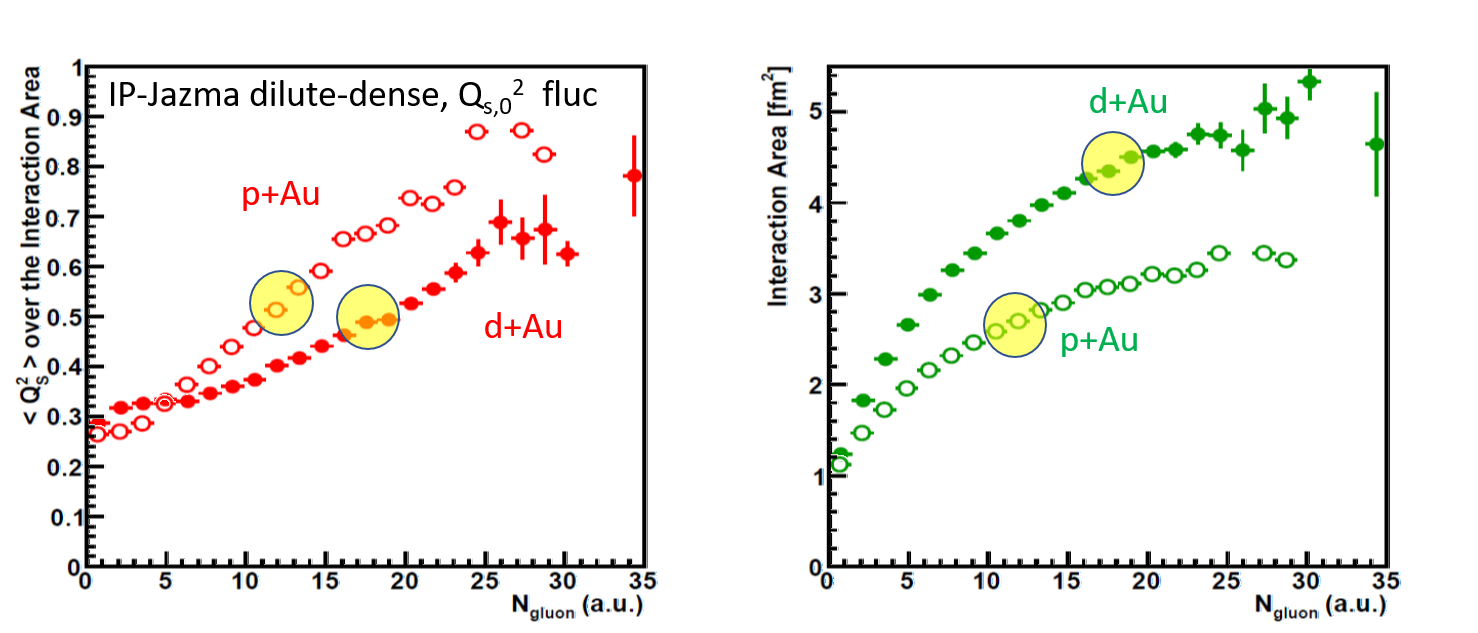}
\caption{\ipjazma calculations of  the interaction area (right) and the average $(Q_s^p)^2)$ for the projectile over that area (left) as a function of the number
of produced gluons $N_{g}$ in arbitrary absolute units.    
While arbitrary, the $x$-axis units are such that one can make a direct comparison between the \pau and \dau results.
The yellow circles indicate the approximate location for the 5\% highest multiplicity selection in both cases, demonstrating that high multiplicity d+Au events result from a larger net interaction area than in p+Au events (RHS), but with similar saturation scales (RHS). 
See text for further discussion.
}
\label{fig:area}
\end{figure*}

In order to further develop our understanding using \ipjazma, we define on an event-by-event basis
the net interaction area by summing the area of all lattice sites with a deposited energy density above some minimum value $\varepsilon_{min}$.   Although $\varepsilon_{min}$ is arbitrary, we apply the identical definition to all
events in both \pau and \dau cases and find that our conclusions are insensitive to this value.   Given 
this well-defined area, one can calculate the average squared saturation scale for the projectile $(Q_s^{proj})^2$ over that area.    
Figure~\ref{fig:area} shows the average area in both \pau and \dau collisions as a function of $N_{g}$, where the x-axis is in arbitrary absolute units but the scaling is common for both \pau and \dau.  The yellow
circles indicate the region in the middle of the 5\% highest multiplicity selection for each collision system.

Quantitatively extracting values for the 5\% highest multiplicity events in both collision systems yields average areas of 2.81 and 4.52~fm$^{2}$ in \pau and \dau respectively.    Similarly, the 
average values for $(Q_s^{proj})^2)$ over those areas are 0.56 and 0.53~GeV$^{2}$ in 
\pau and \dau respectively.    Thus, the multiplicity is 1.5 times higher in high multiplicity \dau compared with \pau (consistent with observations, as noted above) because the area is larger in roughly that same ratio \hbox{$(4.52\ \mathrm{fm}^2)/(2.81\ \mathrm{fm}^2) \approx 1.6$} , while the saturation scale
remains the same.    
This is exactly the opposite of the statement in MSTV,  who argue in a similar comparison of \hau to \pau collisions
that in the dilute-dense limit the higher multiplicity found in 0-5\% \hau collisions 
{\em results from} 
a corresponding increase in saturation scales  $(Q_s^{proj})^2|_{3He} >(Q_s^{proj})^2|_{p} $. In contrast, the \ipjazma
result makes intuitive sense.   If the two nucleons are of order 2~fm apart when striking the
target, they represent two essentially independent proton(neutron)+Au collisions and the saturation scale in the two regions is roughly the same.   
Thus, the
statement in MSTV that in the dilute-dense framework the multiplicity $N_{ch}$ scales 
with $Q_{s}^{2}$ in the projectile is potentially misleading. It is roughly true for a single system (see RHS of 
Figure~\ref{fig:area}), but it is clearly not true {\em across different systems}. Moreover, the dominant source of higher multiplicity events in \dau collisions is increases in the geometric overlap rather than
increases in the saturation scale of the projectile. 

These results and conclusions are consistent with our previous findings. Certainly in the case of 
an infinite target nucleus being struck by two projectile nucleons that are 1 meter apart it is
obvious that the projectile saturation scale will be identical to the case when one nucleon hit the target, but the area is simply twice as large.  
While the Au nucleus is far from infinite, our results presented in Section~\ref{Sec:Overlap} show that
for the loosely bound deuteron, separations of more than 2~fm between the nucleon centers are qualitatively the same as 1 meter, this of course simply being a statement of the confinement scale. 
In the \dau collisions, there of course can be configurations where the two nucleons are one behind the other as they impact the target; in such cases the area will be the same and the projectile saturation scale will be larger.  However, these configurations are quite suppressed by phase space, and in the full \ipjazma calculation for the \dau 5\% highest multiplicity events the average transverse separation is reduced slightly from the unbiased average, but is still greater than 2~fm, consistent with the arguments presented above.  

\section{MSTV Prediction}
\label{Sec:MSTVPrediction}
\begin{figure}[hbtp]
\centering
\includegraphics[width=0.90\linewidth]{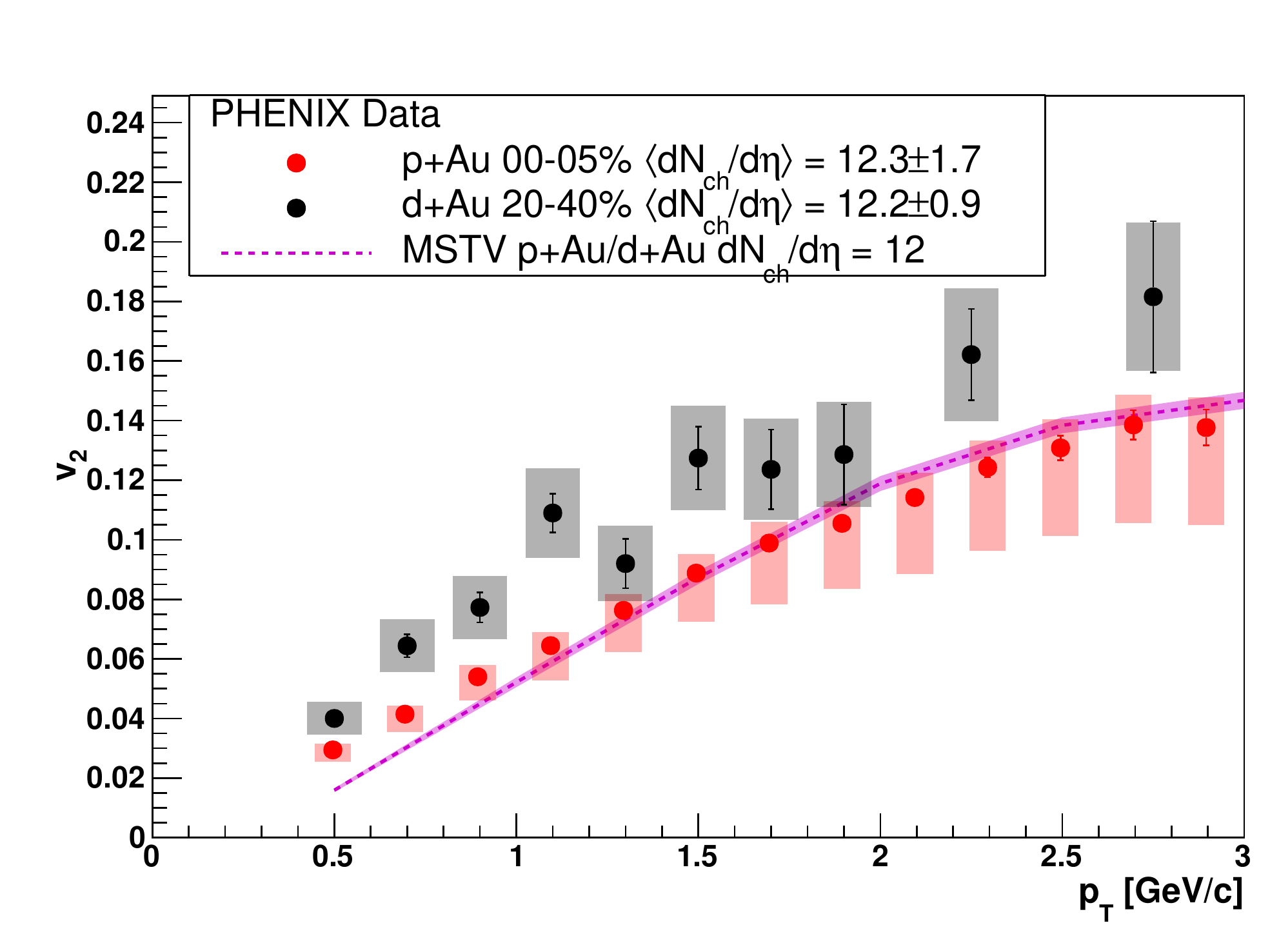}
\caption{PHENIX published data for $v_{2}$ in \pau and \dau collisions in 0-5\% and 20-40\% multiplicity selections, which have comparable average $dN_{ch}/d\eta$ values as shown.    Also shown is the MSTV calculation for \pau at this multiplicity, which should be identical to their result for \dau 20-40\% multiplicity since the $dN_{ch}/d\eta$ is essentially the same.}
\label{fig:phenix}
\end{figure}

Putting aside for the moment the issues elucidated above, there is a prediction explicitly stated in the MSTV paper.  Since the assertion is that the anisotropies for $k_{T} < Q_{s}^{proj}$ are from interactions that are coherent over multiple domains in the projectile, the anisotropies scale with $Q_{s}^{proj}$.    
As noted in the previous section, the authors also state that in the dilute-dense framework the multiplicity $N_{gluon}$ scales with $(Q_s^{proj})^2$.   
As a result,  MSTV predict that if \pau and \dau events are selected with the same $dN_{ch}/d\eta$ then the $v_{2}$ and $v_{3}$ magnitudes and $p_{T}$ dependence should be ``identical".
However, this statement is contradicted by PHENIX data on \dau $v_{2}$ values in different multiplicity classes~\cite{Aidala:2017pup} extant at the time of the MSTV submission. 
The \dau 20-40\% centrality, as defined by the multiplicity in the PHENIX Beam-Beam Counter covering pseudorapidity $-3.9 < \eta < -3.1$ (i.e. in the Au-going direction), has a midrapidity
$dN_{ch}/d\eta = 12.2 \pm 0.9$ and is essentially identical to the \pau 0-5\% centrality with a midrapidity $dN_{ch}/d\eta = 12.3 \pm 1.7$~\cite{Adare:2018toe}.
The comparison of $v_{2}(p_{T})$ values is shown in Figure~\ref{fig:phenix} and highlights that the anisotropies are not the same -- thus contradicting the finding of MSTV.

\section{Other Issues}
\label{Sec:OtherIssues}

The MSTV calculation produces striking agreement of the $v_2(p_T)$ distributions for the 5\% highest centralities in \pau, \dau and \hau collisions, as shown in Figure~4 of their paper. However their calculations are for the $p_T$ of the {\em gluon}, while the data are of course for charged hadrons. 
As shown in Figure~7 of an IP-Glasma calculation of the hadron momentum spectrum~\cite{Schenke:2013dpa} the gluon $p_{T}$ distribution is strikingly different from the hadron $p_{T}$ distribution -- an order of magnitude below it at 0.5~GeV/c and an order of magnitude above at 3.0~GeV/c.   The inclusion of a hadronization scheme is required to roughly reproduce the experimental data in \pp collisions at the LHC.   Its omission in MSTV is notable since other studies with IP-Glasma + PYTHIA include a model dependent version of this hadronization~\cite{Schenke:2018hbz}.
These observations are not original, and in fact are supported by previous statements of a subset of the MSTV authors:
``{\em Fragmentation
of gluons into hadrons will further soften the signal~\cite{Schenke:2016lrs}. Our results} (for $v_n$) {\em therefore represent maximal values
for azimuthal correlations in this initial state framework} \dots {\em Energy evolution of parton distributions and
parton to hadron fragmentation will decrease the values} (of $v_2 (m)(p_\perp)$)
{\em shown.}"~\cite{Dusling:2017dqg}

Another argument that appears in the MSTV paper concerns the relationship between the $k_T$ of a gluon from the target and its ability to resolve color domains in the projectile.  
In the scenario of the dilute-dense limit, MSTV state that those  gluons from the target nucleus satisfying
$k_{T} < Q_{s}^{proj}$ will interact coherently  with 
$ (Q_{s}^{proj} / k_T)^2 $ domains in the projectile.   
As noted previously, we find that the $Q_{s}^{proj}$ values are nearly the same in the 0-5\% highest multiplicity events with deuteron and proton projectiles.
This observation appears to invalidate the MSTV finding of resulting larger $v_{2}$ 
anisotropies in d+Au compared with p+Au.   
While the inclusion of fluctuations in $Q_{s,0}^{2}$ increases the average $Q_{s}^{proj}$ in the 0-5\% highest multiplicity events, 
as seen in Figure~\ref{fig:area}, these values
never approach numbers comparable to 
\hbox{$Q_{s}^{proj} \approx 1.5\mathrm{-}3$~GeV,} 
or equivalently 
\hbox{$(Q_{s}^{proj})^{2} \approx 2.25\mathrm{-}9$~GeV$^{2}$.}
This is a critical observation because the system ordering 
shown in Figure~3 of the MSTV paper indicates that individual domains are not resolved in the projectile all the way up to $p_{T} \approx 2.5\mathrm{-}3.0$~GeV.
That is, even in the presence of fluctuations, the arguments of MSTV appear to require saturation scales in the projectile well in excess of those calculated
with \ipjazma in \pau and \dau collisions at RHIC energies. 

\section{Summary}


We have constructed the \ipjazma model, which provides a very simple implementation of saturation physics phenomenology in the context of Glauber modeling of nuclear collisions. Using this model, we have studied  basic aspects of the dense-dense and dilute-dense frameworks for the CGC in the context of the recent publication by MSTV. We summarize our findings:

\begin{enumerate}

\item The restriction to the 0-5\% centrality bin introduces only a mild bias on the average transverse separation between the neutron and proton on the face of the Au nucleus in \dau collisions (Section~\ref{Sec:MCGlauber}). A quantitative measure of the overlap between the neutron and proton gluon distributions in this centrality class suggests the overlap contribution is at most 20\% and more probably 11\%~(Section~\ref{Sec:Overlap}).

\item \ipjazma, following the simple prescription found in Ref.~\cite{Romatschke:2017ejr}, provides an excellent description of the eccentricity moments $\epsilon_2$ through $\epsilon_6$ over most of the full range of impact parameters in Au+Au collisions, reproducing the results for all but the most peripheral collisions of the full IP-Glasma calculation (Section~\ref{Sec:EnergyDenseDense}). 

\item Using the simplest possible implementation of fluctuations 
in the saturation scale~(Section~\ref{Sec:Fluctuations}),
\hbox{\ipjazma} produces a description 
of the multiplicity distribution in \dau collisions identical to that calculated by MSTV (Section~\ref{Sec:IPJazmaResults}).

\item \ipjazma provides unequivocal support for the intuitive argument that the dominant source of higher multiplicities in \dau collisions is through increases of the interaction area from quasi-independent collisions of the neutron and proton from the deuteron, rather than through local increases in the saturation scale~(Section~\ref{Sec:DandA}). While MSTV do not address this issue directly, their prediction that equal multiplicity \pau and \dau collisions should have identical $v_2(p_T)$ and $v_3(p_T)$ due to the same saturation scale, rests on the underlying assumption that the dominant source of higher multiplicity is via increases in the saturation scale.

\item That same prediction by MSTV is invalidated by existing experimental data for $v_2(p_T)$ in \pau and \dau collisions~(Section~\ref{Sec:MSTVPrediction})

 \item We are unable to reconcile the good agreement MSTV present between $v_2(p_T)$ for {\em gluons} and the PHENIX data for hadrons with the decorrelations both in momentum and angle expected from gluon to hadron fragmentation. In addition, the MSTV argument appears to require saturation scales in \pau and \dau collisions well in excess of those we calculate for these collisions at RHIC energies (Section~\ref{Sec:OtherIssues}).

\end{enumerate}

More generally, 
our \ipjazma calculations suggest that many features attributed to {\em local} color fluctuations and {\em ab initio} features of the CGC are not needed to reproduce multiplicities and eccentricities in nuclear collisions at RHIC. 
In future work we intend to extend \ipjazma to include calculation of momentum spectra and azimuthal anisotropies. 
It may well be that this extension fails, indicating that features intrinsic to the CGC approach are needed for these more microscopic observables. 
Until then, it will not be possible to test the  nature of the coherent correlations over domains and their numerical implementation, including the exact parameter values, with the information available in MSTV.
The proven way to perform scientific assessments of technically involved calculations is through open-source code, as demonstrated by continuing advances in hydrodynamics and jet energy loss. It is our hope that this first \ipjazma work will be useful in that effort.


\section*{ACKNOWLEDGMENTS}

We are pleased to acknowledge very useful discussions with
Jean-Paul Blaizot,
Francois Gelis,
Giuliano Giacalone, 
 Constantin Loizides,
Cyrille Marquet, 
Darren McGlinchey,
Al Mueller,
Bjoern Schenke,
and
Hugo Pereira da Costa, 
We thank    
Tuomas Lappi,
Jean-Yves Ollitraut,
and Paul~Romatschke for a careful reading of the manuscript. 
We also would like to thank 
the MSTV~\cite{Mace:2018vwq} authors
Mark Mace, Vladimir Skokov, Prithwish Tribedy, and Raju Venugopalan
for their detailed descriptions of their calculation 
and their patient answers to our various questions.
JLN and WAZ gratefully acknowledge funding
from the Division of Nuclear Physics of the US Department of
Energy under grants DE-FG02-00ER41152 and DE-FG02-86ER40281,
respectively.  JLN is also thankful for generous support from CEA/IPhT/Saclay during
his sabbatical time in France.

\section{Appendix I}

\begin{figure}[hbtp]
\centering
\includegraphics[width=0.9\linewidth]{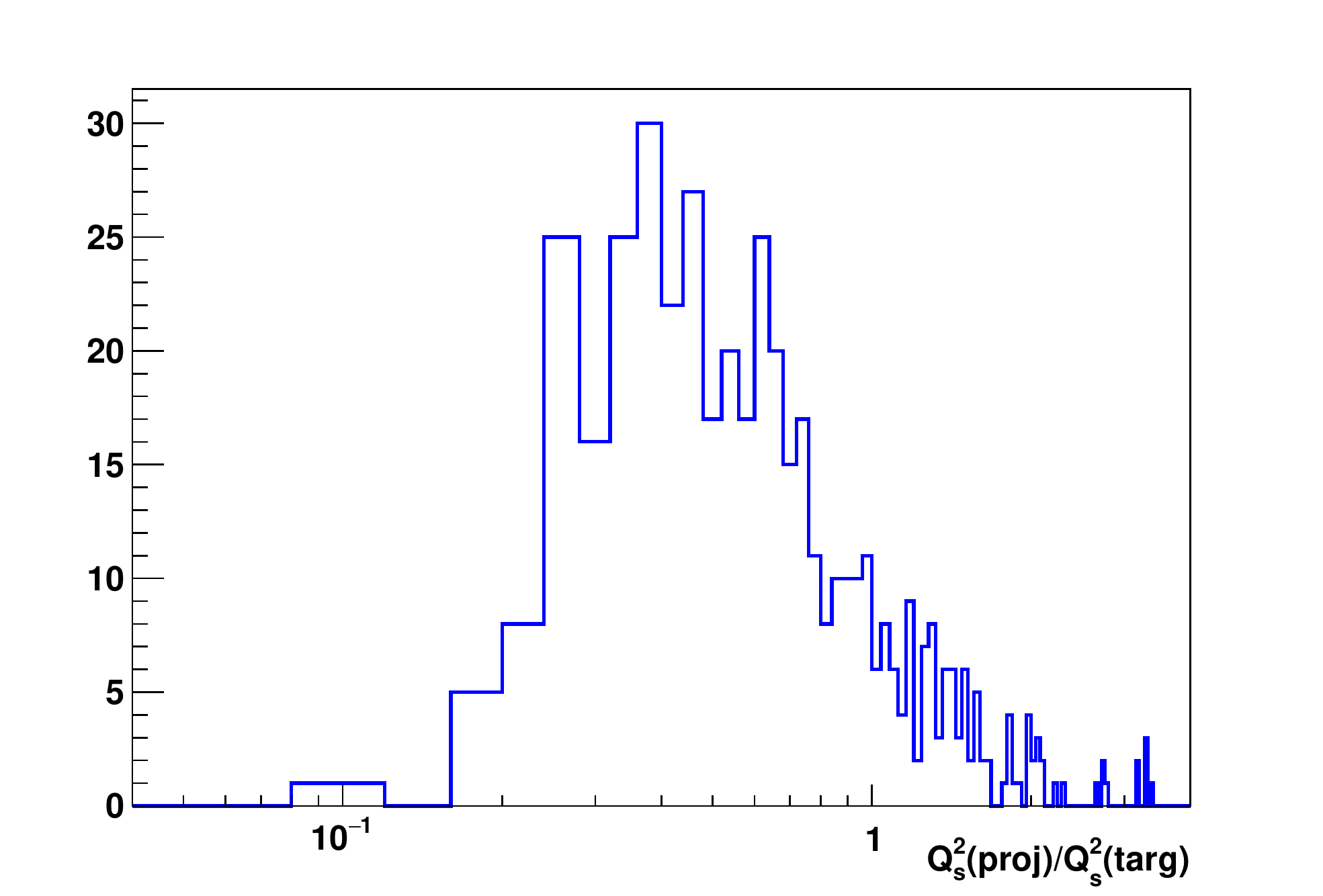}
\caption{\ipjazma \dau highest multiplicity 5\% distribution for $Q_{s}^{2}$(proj) / $Q_{s}^{2}$(targ) weighted by the contribution to the gluon density.}
\label{fig:dilutecheck}
\end{figure}

Here we briefly revisit the question of applying the dense-dense or dilute-dense frameworks.   As mentioned previously, the dilute-dense
framework should be applicable when $Q_{s}$(proj) $< k_{T} < Q_{s}$(targ), which may seem natural in the case of \pau or \dau collisions.
However, a number of IP-Glasma results in the dense-dense limit have been published by some of the same authors and applied also to \pau and \dau collisions at RHIC, as well as to \pp and \ppb collisions at the LHC.  
Given that in MSTV, while working  in the dilute-dense limit,   significant 
fluctuations are added to the saturation scale, it is interesting to check the validity of the condition for applicability of this limit.
For the 5\% highest multiplicity \dau events, we compare the ratio of $Q_{s}^{2}$(proj) / $Q_{s}^{2}$(targ) weighted by the gluon density given in Eqn.~\ref{eqn:dilutedense}.   
The results are shown in Figure~\ref{fig:dilutecheck}.   
The distribution has a peak at $\approx$ 0.4 (corresponding to $\Qsp \sim 0.6\  \Qst\ $), 
but is quite broad, with the mean of the distribution at $\approx$ 0.8, approaching and sometimes exceeding the regime where the saturation scales are equal.
Although the condition for the dilute-dense limit is expressed as a simple inequality, ideally the scales should be well-separated rather than comparable.

\begin{figure}[hbtp]
\centering
\includegraphics[width=0.9\linewidth]{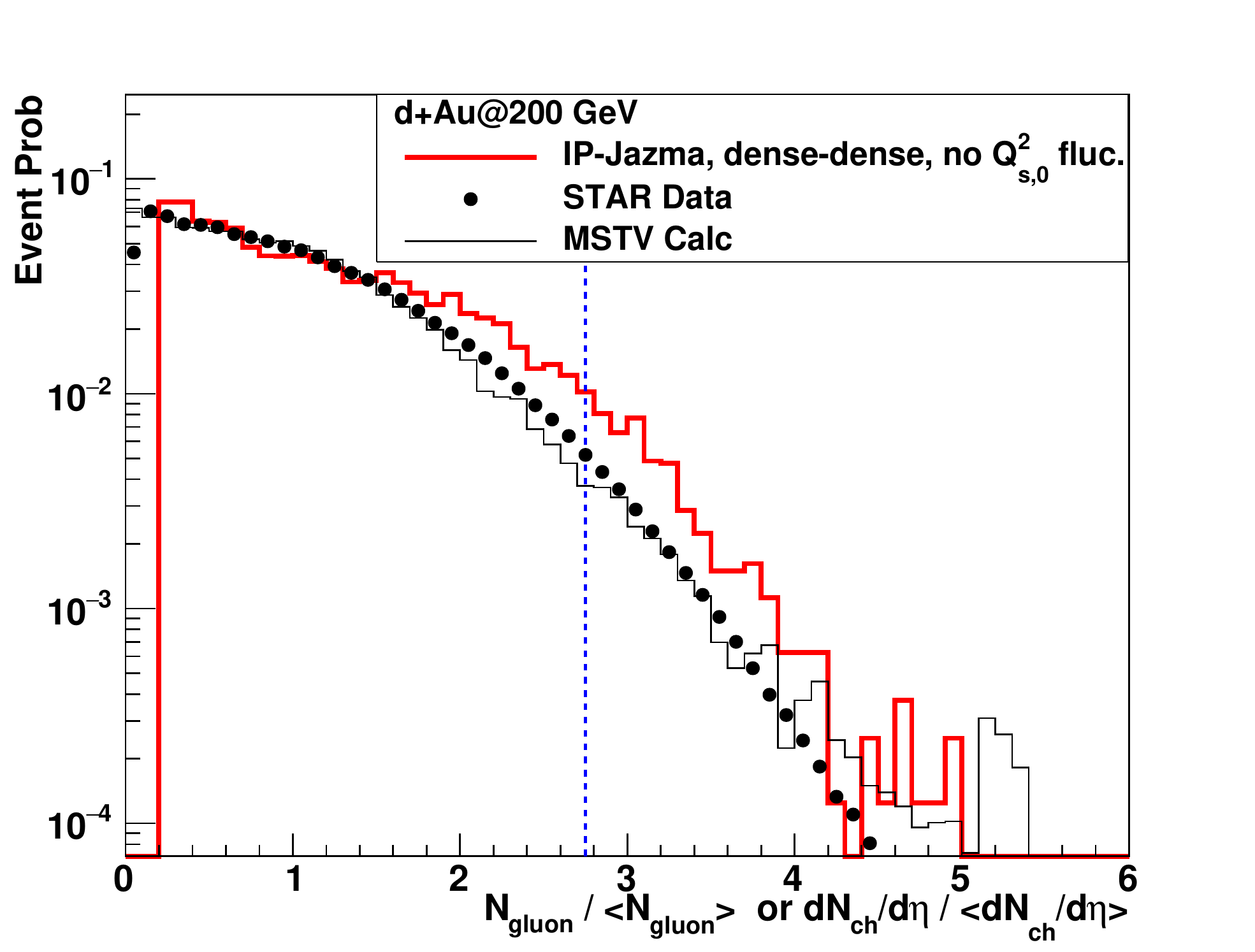}
\caption{\ipjazma \dau minimum bias results for the distribution of $N_{g}/<N_{g}>$ in the dense-dense case and with no $Q_{s,0}^{2}$ fluctuations.}
\label{fig:dau_densedense_nofluc}
\end{figure}

Thus, it is interesting to simply run the \ipjazma calculation for \dau in the dense-dense case, and with
no $Q_{s,0}^{2}$ fluctuations.  
The results are shown in Figure~\ref{fig:dau_densedense_nofluc} and while capturing the overall shape of the data distribution, the agreement is certainly not as good as in the
dilute-dense limit.    Given that no parameters have been tuned, it is plausible that one could achieve a comparable level of agreement as found in the dilute-dense limit.    
This simply demonstrates that one can obtain roughly similar distributions either through fluctuations in the saturation scale of the projectile and a rather flat target or via a constant profile projectile and fluctuations in the nuclear thickness of the target.   
The latter case just validates the fact that the charged particle multiplicity approximately follows constituent quark scaling~\cite{Adare:2015bua}.

\section{Appendix II}

\begin{figure}[hbtp]
\centering
\includegraphics[width=0.9\linewidth]{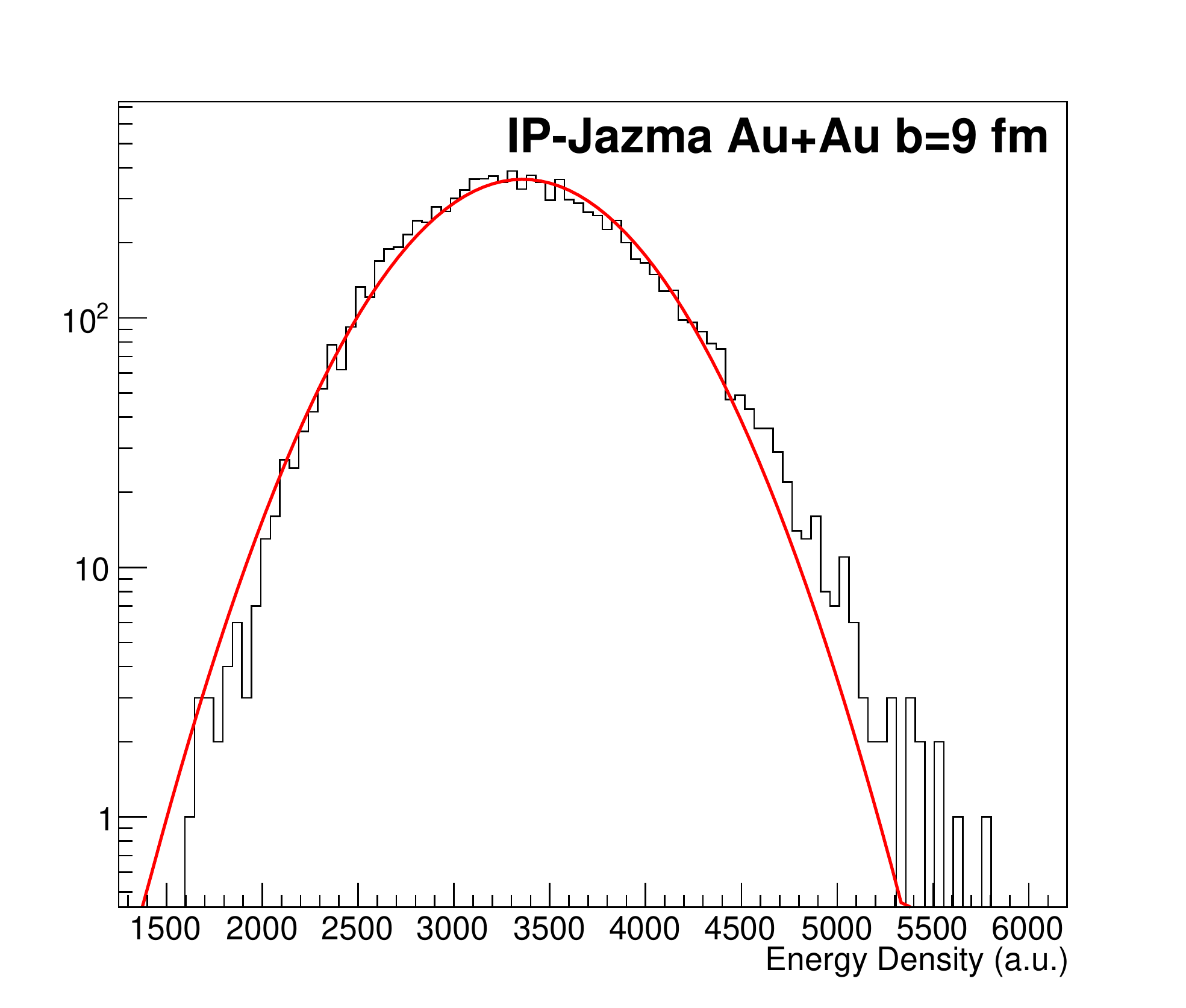}
\caption{\ipjazma \auau b=9 fm events at 200 GeV and their energy density distribution calculated in the dense-dense limit.  Note that no $Q_{s,0}^{2}$ fluctuations are included.   The red curve is a Gaussian fit to the distribution.}
\label{fig:auaub9}
\end{figure}

There is an interesting natural occurrence of negative binomial distribution (NBD) fluctuations within the Color Glass Condensate framework~\cite{Gelis:2009wh} -- referred to as the ``glittering glasma.''   Subsequently, in a number of IP-Glasma and other papers, such NBD behavior is attributed to these gluon field
contributions.    However, there are many sources of fluctuations in the multi-step IP-Glasma and MSTV-type calculations.   As a concrete 
example, Ref~\cite{Schenke:2012wb} uses the IP-Glasma framework to calculate the distribution of transverse energy (proportional to energy density) in a set of
exactly impact parameter $b=9$~fm \auau events at 200~GeV -- see their \hbox{Figure~1}.    The authors find that the distribution is not described by a Gaussian and rather has
a positive skew better described by a NBD.   
(We note here that since energy is a continuous variable, this really should be a Gamma distribution.)
We have studied the same test case using \ipjazma in the dense-dense limit to calculate the distribution of energy density in \auau events with fixed impact parameter $b=9$~fm and plot the event-by-event distribution as shown in Figure~\ref{fig:auaub9}.  The red line is a Gaussian fit to
the distribution, and reveals a clear positive skew in the \ipjazma result.    
Since there are neither fluctuations in $Q_{s,0}^2$ 
nor color or CGC-like fluctuations in this \ipjazma calculation, 
the gamma distribution skew relative to a simple Gaussian must have another source.
That is to say, extreme caution should be used when attributing positive skew in such 
distributions to intrinsic NBD properties of the CGC. 
It is clear from this example that mundane properties of sampling Monte Carlo Glauber configurations with Gaussian profiles produce similar features.

\bibliography{main} 

\end{document}